\documentclass[final]{siamltex}
\usepackage{amssymb}

\usepackage{epsfig}
\usepackage{graphicx}

\begin{document}

\title{Mathematical modeling of pattern formation in sub- and supperdiffusive reaction-diffusion systems}
\author{
Vasyl Gafiychuk\thanks{Institute of Applied Problem of Mechanics and Mathematics, National Academy
of Sciences of Ukraine, Naukova Street 3 B, Lviv, Ukraine 79053(\texttt{viva@iapmm.lviv.ua})}\and
Bohdan Datsko\thanks{Institute of Applied Problem of Mechanics and Mathematics, National Academy
of Sciences of Ukraine, Naukova Street 3 B, Lviv, Ukraine 79053(\texttt{viva@iapmm.lviv.ua})}\and
Vitaliy Meleshko\thanks{Institute of Applied Problem of Mechanics and Mathematics, National Academy
of Sciences of Ukraine, Naukova Street 3 B, Lviv, Ukraine 79053(\texttt{viva@iapmm.lviv.ua})}
}
\maketitle

\begin{abstract}
This paper is concerned with analysis of coupled fractional
reaction-diffusion equations. It provides analytical comparison
for the fractional and regular reaction-diffusion systems. As an
example, the reaction-diffusion model with cubic nonlinearity and
Brusselator model are considered. The detailed linear stability
analysis of the system with cubic nonlinearity is provided. It is
shown that by combining the fractional derivatives index with the
ratio of characteristic times, it is possible to find the marginal
value of the index where the oscillatory instability arises.
Computer simulation and analytical methods are used to analyze
possible solutions for a linearized system. A computer simulation
of the corresponding nonlinear fractional ordinary differential
equations is presented. It is shown that the increase of the
fractional derivative index leads to the periodic solutions which
become stochastic at the index approaching the value of 2. It is
established by computer simulation that there exists a set of
stable spatio-temporal structures of the one-dimensional system
under the Neumann and periodic boundary condition. The
characteristic features of these solutions consist in the
transformation of the steady state dissipative structures to
homogeneous oscillations or space temporary structures at a
certain value of fractional index.
\end{abstract}



\pagestyle{myheadings} \thispagestyle{plain}
\markboth{V.
GAFIYCHUK, B. DATSKO and V. MELESHKO}{V.
GAFIYCHUK, B. DATSKO and V. MELESHKO}

\section{Introduction}

Reaction-diffusion (RD) systems are inherent in many branches of
physics, chemistry, biology, ecology etc. The review of the theory
and applications of reaction-diffusion systems one can find in
books and numerous articles (See, for
example\cite{pr,ch,m90,KO,dk89,dk03,lg,gl,sbg,mo02,g99}). The
popularity of the RD system is driven by the underlying richness
of the nonlinear phenomena, which include stationary and
spatio-temporary dissipative pattern formation, oscillations,
different types of chemical waves, excitability, bistability etc.
The mechanism of the formation of such type of nonlinear phenomena
and the conditions of their emergence have been extensively
studied during the last couple decades. Although the mathematical
theory of such type of phenomena has not been developed yet due to
the essential nonlinearity of these systems, from the viewpoint of
the applied and experimental mathematics, the pattern of possible
phenomena in RD system is more or less understandable.

In the recent years, there has been a great deal of interest in
fractional reaction-diffusion (FRD) systems \cite{hw,hw1,he02,
add,add1,gd,GDI05,ar1,zw,vb} which from one side exhibit
self-organization phenomena and from the other side introduce a
new parameter to these systems, which is a fractional derivative
index, and it gives a greater degree of freedom for diversity of
self-organization phenomena. At the same time, the process of
analyzing such FRD systems is much complicated from the analytical
and numerical point of view.

In this article, we consider two coupled reaction-diffusion systems:

The first one is the classical system

\begin{equation}
\tau _{1}\frac{\partial n_{1}(x,t)}{\partial t}=l^{2}\nabla
^{2}n_{1}(x,t)-W(n_{1},n_{2},\mathcal{A}),  \label{1}
\end{equation}
\begin{equation}
\tau _{2}\frac{\partial n_{2}(x,t)}{\partial t}=L^{2}\nabla
^{2}n_{2}(x,t)-Q(n_{1},n_{2},\mathcal{A}),  \label{2}
\end{equation}
where $x,t\in $ $\Bbb{R}$; $\nabla ^{2}=\frac{\partial ^{2}}{\partial x^{2}}%
;n_{1}(x,t)$, $n_{2}(x,t)\in $ $\Bbb{R}$ -- two variables, $W$ ,$Q\in $ $%
\Bbb{R}$ are the nonlinear sources of the system modeling their production
rates, $\tau _{1},\tau _{2},l,L,\in $ $\Bbb{R}$ -- characteristic times and
lengths of the system, $\mathcal{A}\in $ $\Bbb{R}$ -- is an external
parameter

And the other model is the fractional RD system
\begin{equation}
\tau _{1}\frac{\partial ^{\alpha }n_{1}(x,t)}{\partial t^{\alpha }}%
=l^{2}\nabla ^{2}n_{1}(x,t)-W(n_{1},n_{2},\mathcal{A}),  \label{3}
\end{equation}
\begin{equation}
\tau _{2}\frac{\partial ^{\alpha }n_{2}(x,t)}{\partial t^{\alpha }}%
=L^{2}\nabla ^{2}n_{2}(x,t)-Q(n_{1},n_{2},\mathcal{A})  \label{4}
\end{equation}
with the same parameters and fractional derivatives $\frac{\partial ^{\alpha
}n(x,t)}{\partial t^{\alpha }}$ on the left hand side of equations (\ref{3}%
),(\ref{4}) instead of standard time derivatives, which are the Caputo
fractional derivatives in time of the order $0<\alpha <2$ and are
represented as \cite{skm,po}

\[
\frac{\partial ^{\alpha }}{\partial t^{\alpha }}n(t):=\frac{1}{\Gamma
(m-\alpha )}\int\limits_{0}^{t}\frac{n^{(m)}(\tau )}{(t-\tau )^{\alpha +1-m}}%
d\tau ,\mbox{}\;m-1<\alpha <m,m\in N.
\]

The article is devoted to the second problem and the first one we need for
comparing the obtained results with classical one.

Equations (\ref{3}),(\ref{4}) at $\alpha =1$ correspond to standard RD
system described by equations (\ref{1}),(\ref{2}). At $\alpha <1,$ they
describe anomalous sub-diffusion and at $\alpha >1$~ - anomalous
superdiffusion .

In this paper, we always assume that the following conditions are fulfilled
on the boundaries 0;$l_{x}$:

(i) Neumann: \
\begin{equation}
dn_{i}/dx|_{x=0}=dn_{i}/dx|_{x=l_{x}}=0  \label{bc1}
\end{equation}

(ii) Periodic:
\begin{equation}
n_{i}(t,0)=n_{i}(t,l_{x}),\;dn_{i}/dx|_{x=0}=dn_{i}/dx|_{x=l_{x}},
\label{bc2}
\end{equation}
where $i=1,2.$

\section{Linear stability analysis}

Stability of the steady-state constant solutions of the system (\ref{3}),(%
\ref{4}) correspond to homogeneous equilibrium states
\begin{equation}
W(n_{1},n_{2},\mathcal{A})=0,\;Q(n_{1},n_{2},\mathcal{A})=0
\label{e}
\end{equation}
can be analyzed by linearization of the system nearby this solution. In this
case the system (\ref{3})(\ref{4}) can be transformed to linear system
\begin{equation}
\frac{\partial ^{\alpha }\mathbf{u}(x,t)}{\partial t^{\alpha }}=\widehat{A}%
\mathbf{u}(x,t),  \label{lin}
\end{equation}
where $\mathbf{u}(x,t)=\left(
\begin{array}{c}
\triangle n_{1}(x,t) \\
\triangle n_{2}(x,t)
\end{array}
\right) ,$ $\widehat{A}=\left(
\begin{array}{cc}
(l^{2}\nabla ^{2}-a_{11})/\tau _{1} & -a_{12}/\tau _{1} \\
-a_{21}/\tau _{2} & (L^{2}\nabla ^{2}-a_{22})\tau _{2}
\end{array}
\right) ,$ $a_{11}=W_{n_{1}}^{\prime },$ $a_{12}=W_{n_{2}}^{\prime },$ $%
a_{21}=Q_{n_{1}}^{\prime },$ $a_{22}=Q_{n_{2}}^{\prime }$ (all derivatives
are taken at homogeneous equilibrium states (condition (\ref{e})). By
substituting the solution $\mathbf{u}(x,t)=\left(
\begin{array}{c}
\triangle n_{1}(t) \\
\triangle n_{2}(t)
\end{array}
\right) \cos kx,k=\frac{\pi }{l_{x}}j,j=1,2...$ into FRD system (\ref{3}),(%
\ref{4}) we can get the system of linear ordinary differential equations (%
\ref{lin}) with the matrix $A$ determined by the operator $\widehat{A}$, the
stability conditions of which are given by eigenvalues of this matrix.

Let us analyze the stability of the solution (\ref{e}) of the linear system
with integer derivatives and find the conditions of this instability (See,
for example:\cite{ch,m90,KO,dk89}). We repeat this process in order to
compare the results obtained with the results of the fractional RD system
considered in this article. In this case, by searching for the solution of
the linear system in the form $\mathbf{u}(x,t)=\left(
\begin{array}{c}
\triangle n_{1} \\
\triangle n_{2}
\end{array}
\right) \exp (\lambda t)\cos kx$, we get a homogeneous system of linear
algebraic equations for constants $\triangle n_{1}, \triangle n_{2} $. The
solubility of this system leads to the characteristic equation

\begin{equation}
\det |\lambda I-A|=0,  \label{aa}
\end{equation}
where
\begin{equation}
A=-\left(
\begin{array}{cc}
(l^{2}k^{2}+a_{11})/\tau _{1} & a_{12}/\tau _{1} \\
a_{21}/\tau _{2} & (L^{2}k^{2}+a_{22})/\tau _{2}
\end{array}
\right),  \label{a}
\end{equation}
$I$ \ is the identity matrix. As a result, the characteristic equation takes
on a form of a simple quadratic equation $\lambda ^{2}-trA\lambda +\det A=0.$%
\

The linear boundary value problem for RD system (\ref{1}),(\ref{2}) is
unstable according to inhomogeneous wave vectors $k\neq 0$ if ($trA<0,\det
A<0$)
\begin{equation}
a_{11}<-[(l/L)^{2}a_{22}+2(l/L)(a_{22}a_{11}-a_{12}a_{21})^{1/2}],
\label{tu}
\end{equation}
\[
\tau _{2}(a_{11}+k^{2}l^{2})+\tau
_{1}(a_{22}+k^{2}L^{2})>0,\,a_{22}a_{11}-a_{12}a_{21}>0
\]
(Turing Bifurcation) and according to homogenous ($k=0$) fluctuations (Hopf
Bifurcation) if ($trA>0,\det A>0$)
\begin{equation}
\tau _{2}a_{11}+\tau _{1}a_{22}<0,a_{22}a_{11}-a_{12}a_{21}>0.  \label{h}
\end{equation}

For analyzing the equations (\ref{3}),(\ref{4}) let us also consider a linear
system obtained near the homogeneous state (\ref{e}). As a result, the
simple linear transformation can convert this linear system to a diagonal
form

\begin{equation}
\frac{d^{\alpha }\mathbf{\eta (}t\mathbf{)}}{dt^{\alpha
}}=C\mathbf{\eta }(t) \label{f},
\end{equation}
where $C$ is a diagonal matrix of $A$: $\ C=P^{-1}AP=\left(
\begin{array}{cc}
\lambda _{1} & 0 \\
0 & \lambda _{2}
\end{array}
\right) ,$ eigenvalues $\lambda _{1,2}$ are determined by the same
characteristic equation (\ref{aa}) with matrix (\ref{a}), $\lambda _{1,2}=%
\frac{1}{2}(trA\pm \sqrt{tr^{2}A-4\det A}),\ \mathbf{\eta }(t)=P^{-1}\left(
\begin{array}{c}
\triangle n_{1}(t) \\
\triangle n_{2}(t)
\end{array}
\right) ,$ $P$ \ is the matrix of eigenvectors of matrix $A$.

\begin{figure}[tbp]
\begin{center}
\begin{tabular}{cc}
\includegraphics[width=0.35\textwidth]{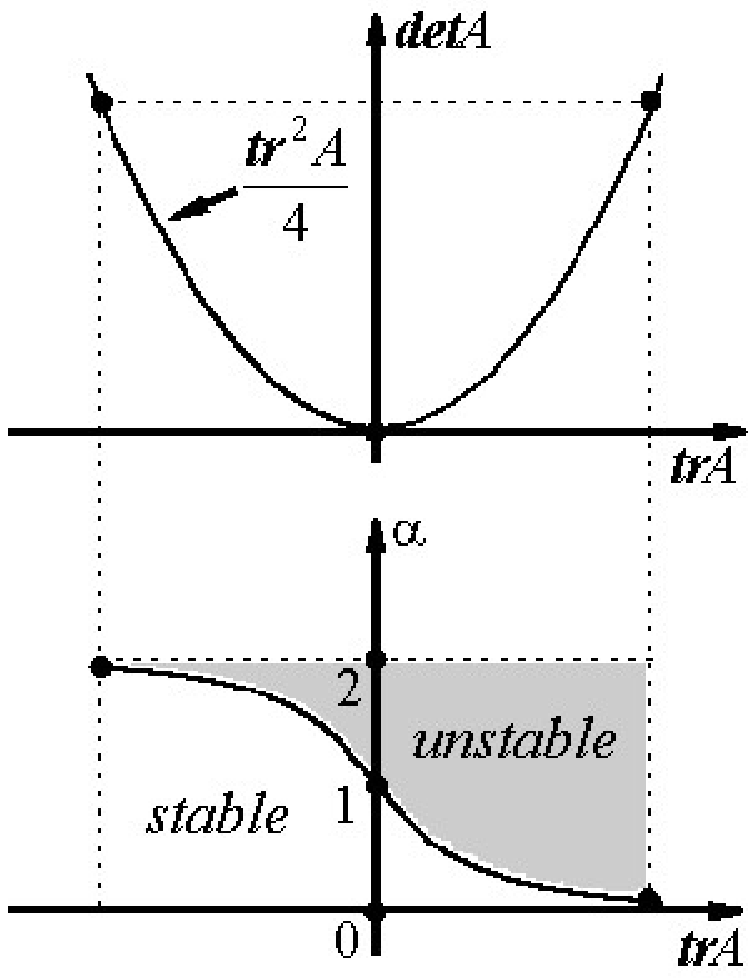} & %
\includegraphics[width=0.47\textwidth]{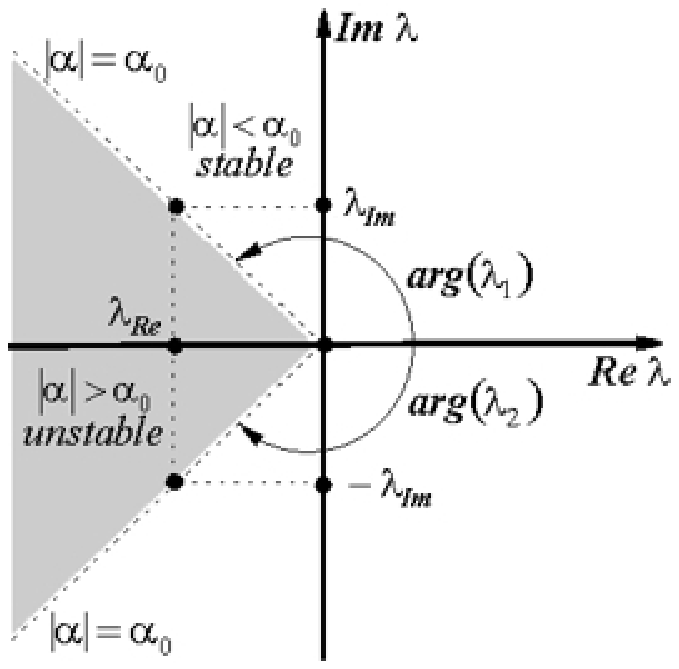} \\
(a) & (b)
\end{tabular}
\end{center}
\caption{Schematic view of the marginal curve describing fixed points for
two-dimensional vector field -- (a) the marginal value of $\protect\alpha$ --
(b).}
\label{rys1}
\end{figure}

In this case, the solution of the vector equation (\ref{f}) is given by
Mittag-Leffler functions\bigskip\ \cite{skm,po,os06,a,tz}

\begin{equation}
\triangle n_{i}(t)=\sum\limits_{k=0}^{\infty }\frac{(\lambda _{i}t^{\alpha
})^{k}}{\Gamma (k\alpha +1)}\triangle n_{i}(0)=E_{\alpha }(\lambda
_{i}t^{\alpha })\triangle n_{i}(0), \quad i=1,2.  \label{sol}
\end{equation}
Using the result obtained in the papers \cite{GDI05,mi}, we can
conclude that if for any of the roots
\begin{equation}
|arg(\lambda _{i})|<\alpha \pi /2  \label{100}
\end{equation}
the solution has an increasing function component then the
system is asymptotically unstable.

Analyzing the roots of the characteristic equations, we can see that at $%
4\det A-tr^{2}A>0$ eignevalues are complex and can be represented as
\[
\lambda _{1,2}=\frac{1}{2}(trA\pm i\sqrt{4\det A-tr^{2}A})\equiv
\lambda _{Re}\pm i\lambda _{Im}.
\]
The roots $\lambda _{1,2}$ are complex inside the parabola (Figure \ref{rys1}%
(a)) and the fixed points are the spiral source ($trA>0$)\ \ or spiral sinks (%
$trA<0$). The plot of the marginal value $\alpha :\alpha =\alpha _{0}=$ $%
\frac{2}{\pi }|arg(\lambda _{i})|$ which follows from the conditions (\ref
{100}) is given by\ the formula

\begin{equation}
\alpha _{0}=\left\{
\begin{array}{cc}
\frac{2}{\pi }\arctan \sqrt{4\det A/tr^{2}A-1}, & trA\geq 0, \\
2-\frac{2}{\pi }\arctan \sqrt{4\det A/tr^{2}A-1}, & trA\leq 0
\end{array}
\right.  \label{al}
\end{equation}
and is presented in the Figure \ref{rys1}(a) below the parabola in the
coordinate system $(trA,\alpha ).$

Let us analyze the system solution with the help of the Figure
\ref{rys1}(a). Consider the parameters which keep the system
inside the parabola. It is a well-known fact, that at $\alpha =1$\
the domain on the righthand side of the parabola ($trA>0$) is
unstable with the existing limit circle, while the
domain on the left hand side ($trA<0$) is stable. By crossing the axis $%
trA=0 $ the Hopf bifurcation conditions become true. In the general case of
\ $\alpha :0<\alpha < 2$\ \, for every point inside the parabola there
exists a marginal value of $\alpha _{0}$ where the system changes its
stability. The value of $\alpha $ is a certain bifurcation parameter which
switches the stable and unstable state of the system. At lower $\alpha :$ $%
\alpha <\alpha _{0}=$ $\frac{2}{\pi }|arg(\lambda _{i})|$ the system has
oscillatory modes but they are stable. Increasing the value of $\alpha
>\alpha _{0}=$ $\frac{2}{\pi }|arg(\lambda _{i})|$ leads to instability. As
a result, the domain below the curve\ $\alpha _{0}$, as a function of $trA$%
\, is stable and the domain above the curve is unstable.

The plot of the roots, describing the mechanism of the system
instability, can be understood from the Figure \ref{rys1}(b) where
the case $\alpha_{0}>1$ is described. In fact, having complex
number $\lambda_{i}$ with ${\it Re} \lambda_{i}<0$\ at $\alpha \rightarrow 2$
it is always possible to satisfy the condition
$|arg(\lambda_{i})|<\alpha \pi /2$, and the system becomes
unstable according to homogeneous oscillations (Figure
\ref{rys1}(b)). The smaller is the value of $trA$, the easier it is
to fulfill the instability conditions.

In contrast to this case, a complex values of $\lambda _{i},$
with$\ {\it Re} \lambda _{i}>0$ lead to the system instability for regular system with $%
\alpha=1 .$ However fractional derivatives with $\alpha <1$ will
stabilize the system if $\alpha <\alpha _{0}=$ $\frac{2}{\pi
}|arg(\lambda _{i})|.$ This
makes it possible to conclude that fractional derivative equations with $%
\alpha <1$\ are more stable that their integer twinges.

\section{Solutions of the coupled fractional ordinary differential equations
(FODEs)}

Our particular interest here is the analysis of the specific
non-linear system of FRD equations. We consider two very
well-known examples. The first one is the RD system with cubical
nonlinearity \cite{m90,KO,lg} which probably is the simplest one
used in RD systems modeling
\begin{equation}
W(n_{1},n_{2})=-n_{1}+n_{1}^{3}/3+n_{2},\quad Q=n_{2}-\beta n_{1}-\mathcal{A}%
.  \label{nlin2}
\end{equation}
The second example is known as Brusselator model \cite{pr} and it describes
certain chemical reaction-diffusion processes with a pair of variables whose
concentrations are controlled by nonlinearities
\begin{equation}
W(n_{1},n_{2})=-\mathcal{A}+(\beta +1)n_{1}-n_{1}^{2}n_{2},\quad
Q(n_{1},n_{2})=- \beta n_{1}+n_{1}^{2}n_{2},  \label{nlin1}
\end{equation}

\begin{figure}[tbp]
\begin{center}
\begin{tabular}{cc}
\includegraphics[width=0.9\textwidth]{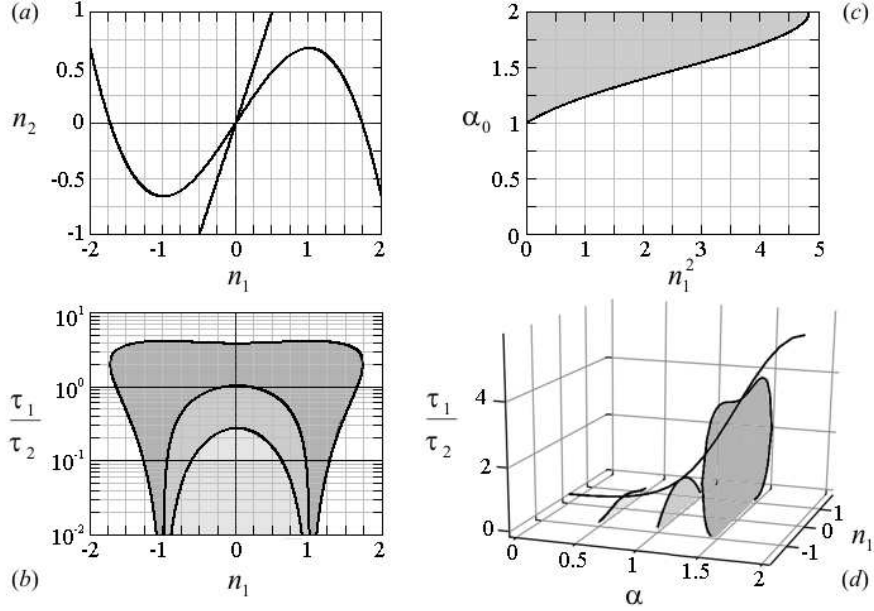} &
\end{tabular}
\end{center}
\caption{Null isoclines -- (a), Instability domains (shaded regions) in coordinates $(\overline{n}%
_{1},\protect\tau _{1}/\protect\tau _{2})$ (Dependence of
$\protect\tau _{1}/\protect\tau _{2}$ on $\overline{n}_{1}$) -- (b), Dependence of $%
\protect\alpha _{0}$ on $\overline{n}_{1}^{2}$ -- (c), Dependence of $\protect%
\tau _{1}/\protect\tau _{2}$ on $\protect\alpha _{0}$ (the domains
in coordinates $(\overline{n}_{1}$, $\protect\tau
_{1}/\protect\tau _{2}$) correspond to domains represented on
figure (b) -- (d). } \label{rys2}
\end{figure}

Let us first consider the coupled fractional ordinary differential
equations (FODEs) with nonlinearities (\ref{nlin2}) and analyze
the stability conditions for such systems. The plot of isoclines
for the system (\ref {nlin2}) is represented on Figure \ref{rys2}(a).
In this case, for$\ $
homogeneous solution, which can be determined from the system of equations $%
W=Q=0$, is the solution of cubic algebraic equation\ \ ($\beta -1)\overline{n%
}_{1}+\overline{n}_{1}^{3}/3+\mathcal{A}=0.$ Simple calculation makes it
possible to write the expressions required for analysis $A=-\left(
\begin{array}{cc}
(-1+\overline{n}_{1}^{2})/\tau _{1} & \quad 1/\tau _{1} \\
-\beta /\tau _{2} & \quad 1/\tau _{2}
\end{array}
\right) ,$ $trA=(1-\overline{n}_{1}^{2})/\tau _{1}-1/\tau _{2},$ $\quad \det
A= ( (\beta -1)+\overline{n}_{1}^{2})/\tau _{1}\tau _{2} .$

It is easy to see that if the value of $\tau _{1}/\tau _{2}$, in
certain cases, is smaller than $1$, the instability conditions (
$trA>0$)\ lead to Hopf bifurcation for regular system ($\alpha
=1$) \cite{pr,ch,m90,KO,dk89}. In this case, the plot of the
domain, where instability exists, is shown on the Figure
\ref{rys2}(b).

The linear analysis of the system for $\alpha =1$ shows that, if
$\tau _{1}/\tau _{2}>1$, the solution corresponds to the
intersections of two isoclines, and it is stable.  The marginal
curve, separating stability and instability domains, is given on
Figure \ref{rys2}(b). The smaller is the ratio of $\tau _{1}/\tau
_{2}$, the wider is the instability region.
Formally, at $\tau _{1}/\tau _{2}\rightarrow 0$, the instability region in $%
\overline{n}_{1}$\ coinsides with the interval ($-1,1$) where the
null isocline $W(n_{1},n_{2})=0$ \ has its increasing part. The
maximum value of the curve $\tau _{1}/\tau _{2}(n_1)$ corresponds
to the value $\tau _{1}/\tau _{2}=1$ where the system is neutrally
stable. These results are very widely known in the theory of
nonlinear dynamical systems \cite{pr,ch,m90,KO,dk89}.

In the FODEs the conditions of the instability change (\ref{100}), and we
have to analyze the real and the imaginary part of the existing complex
eigenvalues, especially the equation: $4\det A-tr^{2}A=4((\beta -1)+%
\overline{n}_{1}^{2})/\tau _{1}\tau _{2}-\left( (1-\overline{n}%
_{1}^{2})/\tau _{1}-1/\tau _{2}\right) ^{2}>0.$ In fact, with the
complex eigenvalues, it is possible to find out the corresponding
value of $\alpha $ where the condition (\ref{100}) is true. We
show that this interval is not correlated with the increasing part
of the null isocline of the system. Indeed, omitting simple
calculation, we can write an equation for marginal
values of $\overline{n}_{1}$: $\overline{n}_{1}^{4}-2(1+%
\frac{\tau _{1}}{\tau _{2}})\overline{n}_{1}^{2}+\frac{\tau _{1}^{2}}{\tau
_{2}^{2}}-2\frac{\tau _{1}}{\tau _{2}}(2\beta -1)+1=0,$ and expression $%
\overline{n}_{1}^{2}=1+\frac{\tau _{1}}{\tau _{2}}\pm 2\sqrt{\beta \frac{%
\tau _{1}}{\tau _{2}}}$ estimates the maximum and minimum values of $%
\overline{n}_{1}$ where the system can be unstable at certain value of $%
\alpha =\alpha _{0}$.  For example, examine the domain of the FODEs where
the eigenvalues are complex for fixed value $\tau _{1}/\tau _{2},$ for
example, consider $\tau _{1}=\tau _{2}=1$ and $\beta =2.$ In this case, $%
trA=-\overline{n}_{1}^{2},\det A=1+\overline{n}_{1}^{2},4\det A-tr^{2}A=4+4%
\overline{n}_{1}^{2}-\overline{n}_{1}^{4}>0,$ which immediately leads to the
region of existing of complex roots $-\sqrt{2+2\sqrt{2}}\leq \overline{n}%
_{1}\leq \sqrt{2+2\sqrt{2}}$ and we can conclude that the instability region, due to the
fractional order of the derivatives, can be much wider
than the same region for ($\alpha =1)$. The dependance of the value of $%
\alpha $ on the value $\overline{n}_{1}^{2}$ (\ref{al})\ is given
on Figure \ref{rys2}(c).

In this case the plot is obtained at $\tau _{1}/\tau _{2}=1$ and
that is why the marginal instability curve is determined for
$\alpha >1$ (for $\alpha <1$ the system is stable). On this figure
the domain below the curve corresponds to stability \ and above it
to instability conditions.

Similar analysis can be provided for $trA>0$ where $(\tau
_{1}/\tau _{2})$ is smaller than unity. In this case, the
instability conditions are also not correlated with the increasing
part of the null isocline $W_{1}(n_{1},n_{2}).$ In this case the
plot of  $\alpha $ will be start not from 1 but from certain value
smaller than unity. However, it is much better for understanding
to get marginal curve $\alpha _{0}$ as a function of  $\tau
_{1}/\tau _{2}$.

Let us analyze the dependence of $\alpha _{0}$ on the parameter
$\tau _{1}/\tau _{2}$ where the system changes its stability. As
it is easy to see from Figure
\ref{rys2}(a),(b)\ that the easiest way to reach instability domain is realized at $%
\mathcal{A}=0$\ when two isoclines are intersected themselves at
the point $(0,0) $ and this corresponds to maximum of the curve on
Figure \ref{rys2}(b)\ . Let us consider the dependence of this\
marginal value of $\alpha _{0}$ on the parameter $\tau _{1}/\tau
_{2}$ at $\mathcal{A}=0$ and determine the plot of this maximum \
as a functions of $\alpha _{0}$ \ on $\tau _{1}/\tau _{2}.$\ Such
curve  for the given model is represented in the Figure
\ref{rys2}(d). This curve obtained from (\ref{al}) corresponds to
the dependence of $\alpha $ on $trA$ (Figure \ref{rys1}(a)). Below
this curve the system is unstable, and above it - it is stable. We
may therefore focus our attention on the
general form of this curve. At sufficiently small value $\tau _{1}/\tau _{2}$%
\ oscillatory instability is valid even for small $\alpha <1$. In contrast,
at $\alpha >1$, the instability conditions could have place even for those
cases when $\tau _{1}/\tau _{2}$ is sufficiently large. This means that
fractional differential equations, by corresponding combination of the
parameter $\tau _{1}/\tau _{2}$, can be stable or unstable practically in
all the region $1<\alpha <2.$

It should be noted that, even if the eigenvalues are not complex
($\lambda _{Im}=0$), the systems with fractional derivatives can
poses oscillatory damping oscillations. Such situation takes
place when $4\det A-tr^{2}A<0,$ $trA<0,$ $\det A>0$ and two
eigenvalues are real and less than zero. In this case, at
$1<\alpha <2$\ steady state solutions of the system are stable and
any perturbations are damping. Such system was considered, for
example, in the article \cite{zse05},\ where an analytical
solution for fractional oscillator is obtained. Namely with this
case we start analyzing possible solutions.

Several examples of linear FDOEs were solved analytically by Adomnian
decomposition method, as well as numerically. The obtained solutions were
compared with the analytical solutions obtained by Mittag-Leffler functions (%
\ref{sol}) and the results of the work \cite{zse05}.

\begin{figure}[tbp]
\begin{center}
\begin{tabular}{cc}
\includegraphics[width=0.49\textwidth]{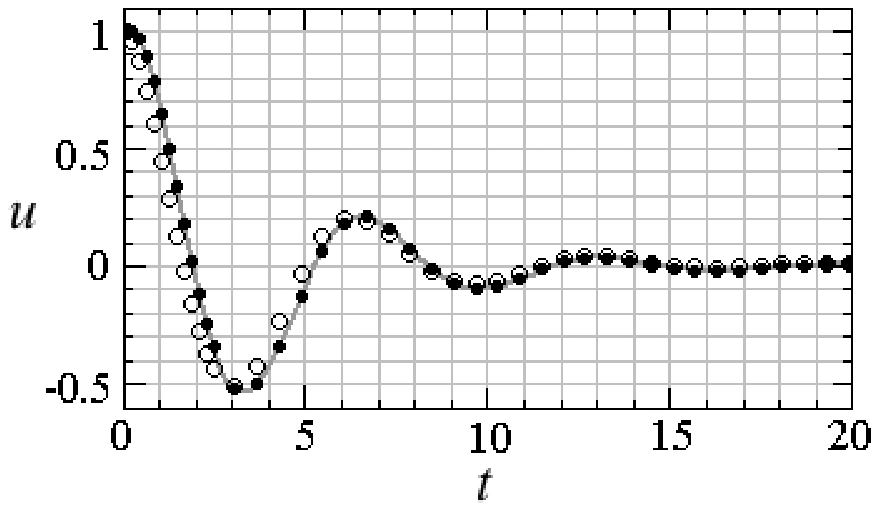} & %
\includegraphics[width=0.5\textwidth]{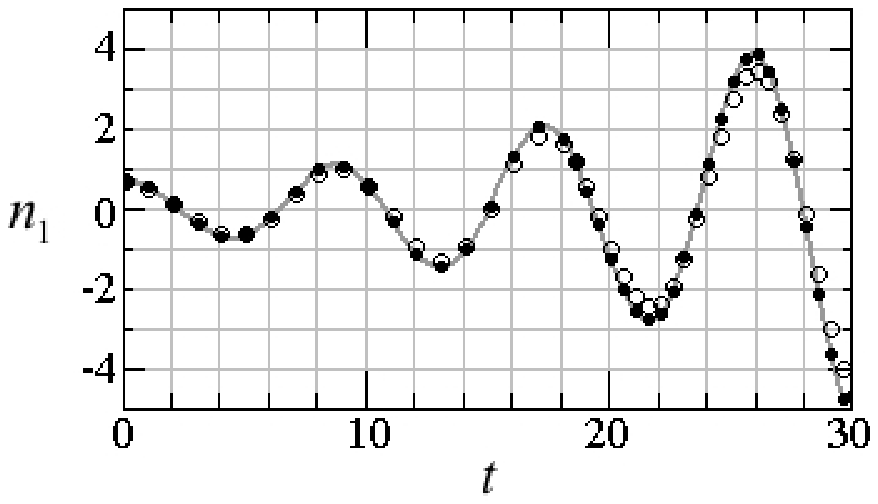} \\
(a) & (b)
\end{tabular}
\end{center}
\caption{The damped oscillator solution for $\frac{\partial^\alpha
u}{\partial t^\alpha}=-u \quad at \quad \protect\alpha=1.7, \quad
u(0)=1$ -- (a), and the increasing time domain oscillations of the
linearized system (\ref{3}),(\ref{4}) for $n_1$ and
$\protect\alpha =1.3, \mathcal{A}=-0.1, \beta=1, \tau _{1}=\tau
_{2}=1, n_1(0)=(-3\mathcal{A})^{\frac{1}{3}},
n_2(0)=(-3\mathcal{A})^{\frac{1}{3}}+\mathcal{A}$ -- (b). Small
filled circled  line -- Mittag-Leffler solution, empty circled
line -- Adomnian decomposition  solution, solid gray line --
numerical solution } \label{rys3}
\end{figure}

Three different solutions are plotted on the same Figure
\ref{rys3}. The results of the computer simulation of linear
ordinary differential equations for stable system - a) and
unstable system - b) of one variable $n_{1}$ are given on the time
domain interval $[0,30]$. We can see that in the stable domain
$\alpha <\alpha _{0}$ the oscillations are damping, and at $\alpha
>\alpha _{0}$ they grow exponentially. So, the ordinary
differential equations (space derivatives are equal to zero) of
the system have two different modes. The solution is either
asymptotically stable or unstable.

It is a statement, that FODEs are at least as stable as their integer order
counterparts \cite{po}. At the same time, we have established here that the
dynamics of the FODEs can be much more complicated than that of the
equations with integer order \cite{El96,mi,wz}. Our task here is to clarify
these statements by finding out not only the conditions of the bifurcation
but also the real time dynamics of FODEs.

The developed technique of Adomian decomposition \cite{a94,bbi04,jd06,db04}
is a powerful method for solving the system of ordinary differential
equations. The effectiveness of this method was demonstrated for solving
linear fractional differential equations, and we used this method here to
test the computer program and to compare these analytical results with
Mittag-Leffler functions.

First, we consider the case when the system is always asymptotically stable
at $\alpha <\alpha _{0}=2$, for example $\alpha=1.7$  (Figure \ref{rys3}(a)).
Three solutions obtained by these different methods practically do not differ from
each other and they show oscillatory damping time behavior (the model described in \cite{zse05}).
Analytical results obtained by Adomian decomposition is given by the formula
\begin{equation}
n_{1}=\sum_{k=0}^{\infty}(-1)^{k}\frac{t^{k\alpha}}{\Gamma (k\alpha +1)},
\end{equation}
which is plot on the Figure \ref{rys3}(a) by empty circled curve
(we truncate the series at k=60). At small time interval (for
example $t<7$) this series can be represented by next expansion
\begin{eqnarray*}
n_{1} &=& 1-0.6473808267\cdot t^{1.7}+0.9865725648\cdot 10^{-1}\cdot
t^{3.4}-0.7019911214\cdot 10^{-2}\cdot t^{5.1}
\\&&+0.296127716\cdot 10^{-3}\cdot t^{6.8}-0.838275933\cdot 10^{-5}\cdot
t^{8.5}+ 0.171848173\cdot 10^{-6}\cdot t^{10.2}
\\&&-0.2686528893\cdot 10^{-8}\cdot t^{11.9}+ 0.3324725330\cdot 10^{-10}\cdot t^{13.6}
\\&&-0.3350625811\cdot 10^{-12}\cdot t^{15.3}+0.2811457252\cdot 10^{-14}\cdot t^{17}.
\end{eqnarray*}
Despite the considered system, in this particular case, cannot be
unstable (Figure \ref{rys3}(a)) because the values of $\lambda $
are real and negative $|arg(\lambda _{1})|=\pi >\alpha \pi /2$, in
the limit at $\alpha =2$ we have a regular linear oscillator
\cite{zse05}, analytical solution of which we get from Adomian
decomposition in the form of power series. The terms of this
series completely coincides with the expansion of linear
oscillator solution $cos(x)$.

It should be noted that all damped oscillations solutions of the
linear oscillator of fractional order have similar plots.

Quite different is the dynamics of FODEs for $\alpha >\alpha _{0}$ when
oscillatory mode becomes unstable and leads to increasing oscillations at $%
\alpha >\alpha _{0}$ (Figure \ref{rys3}(b)). In this case, the
analytical solution obtained by Adomian decomposition for
$\alpha =1.3$ looks  like
\begin{eqnarray*}
n_{1} &=&0.67-0.1716397314\cdot t^{1.3}-0.05663459553\cdot
t^{2.6}+0.008906834532\cdot t^{3.9} \\
&&+0.00006991997714\cdot t^{5.2}-0.00004692163249\cdot
t^{6.5}+0.1292316598\cdot 10^{-5}\cdot t^{7.8} \\
&&+0.5298004324\cdot 10^{-7}\cdot t^{9.1}-0.2781417477\cdot 10^{-8}\cdot t^{10.4}
\\
&&+0.3895105489\cdot 10^{-11}\cdot t^{11.7}+0.1811701048\cdot 10^{-11}\cdot
t^{13}+...
\end{eqnarray*}

The result of taking into account 50 terms in Adomian
decomposition expansion makes it possible to represent the
solution in the time interval till $t=30$ is presented on the
Figure \ref{rys3}(b). As a matter of fact, such representation has
rather theoretical sense because our system is essentially
nonlinear and does not allow this type of solutions. For the
initial stage dynamics or for the linear system, Adomian
decomposition method is very effective. The application of Adomian
decomposition to the nonlinear FDOEs is not successful. Taking
into account 10 terms makes it possible to find out the solution
on time interval $t=6-7$ and at higher values of $t$ the
discrepancy between the numerical solution and the one, obtained
by Adomian decomposition, increases rapidly.

To demonstrate the nontrivial properties of FDOEs, here we consider the
nonlinear dynamics of the above mentioned nonlinear fractional differential
equations.

\begin{figure}[tbp]
\begin{center}
\begin{tabular}{cc}
\includegraphics[width=0.35\textwidth]{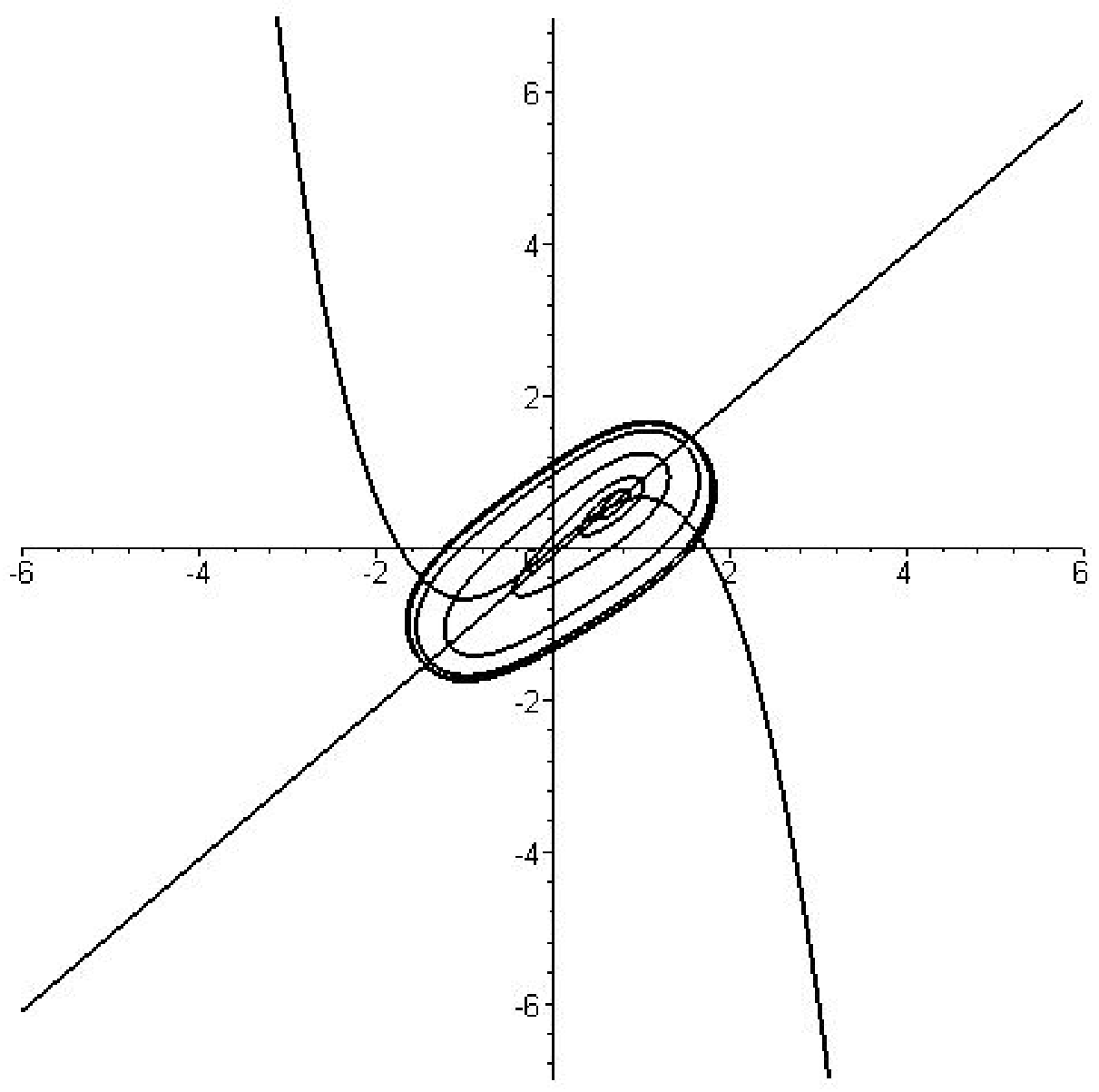} & %
\includegraphics[width=0.35\textwidth]{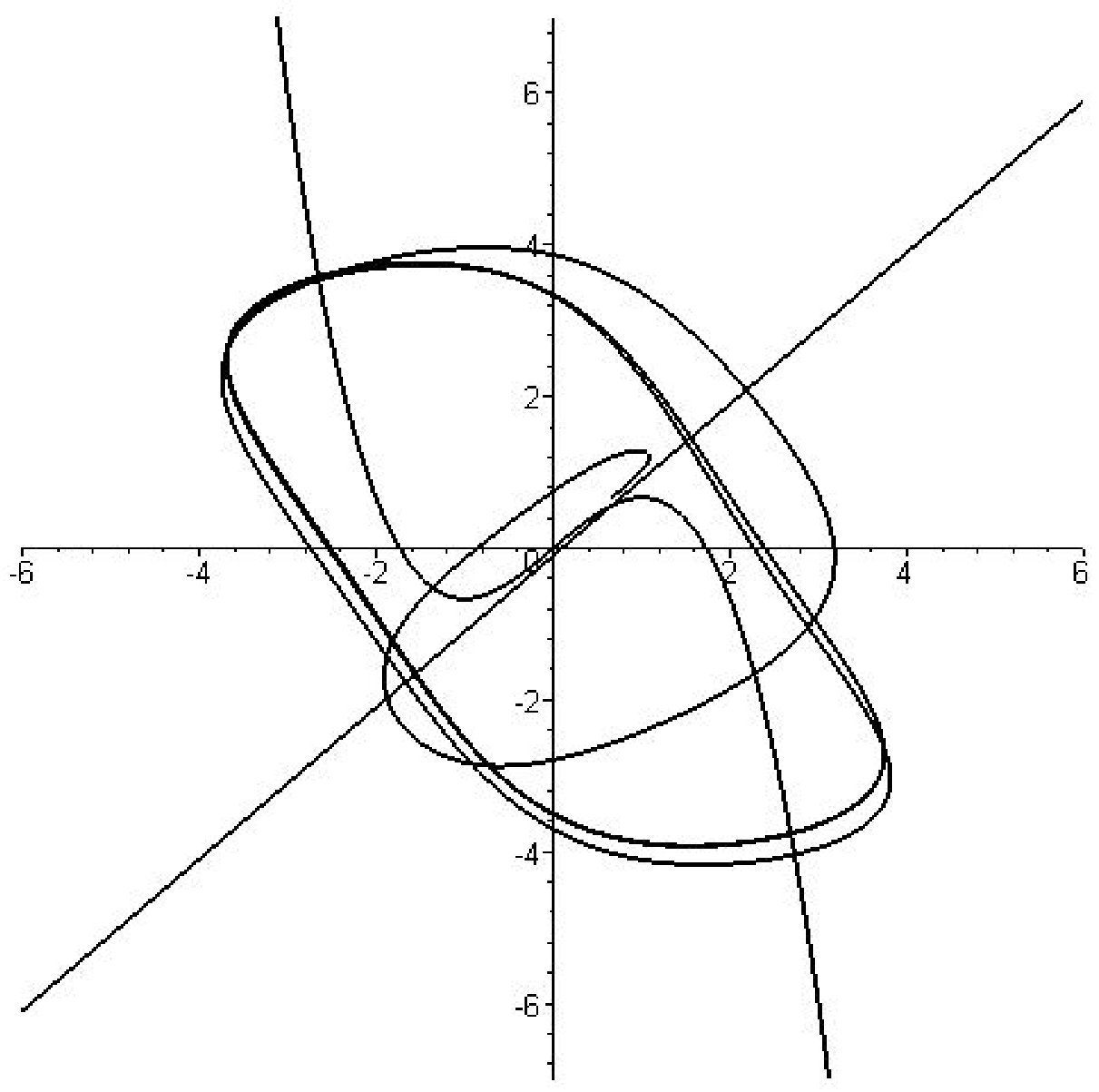} \\
(a) & (b)
\end{tabular}
\begin{tabular}{cc}
\includegraphics[width=0.35\textwidth]{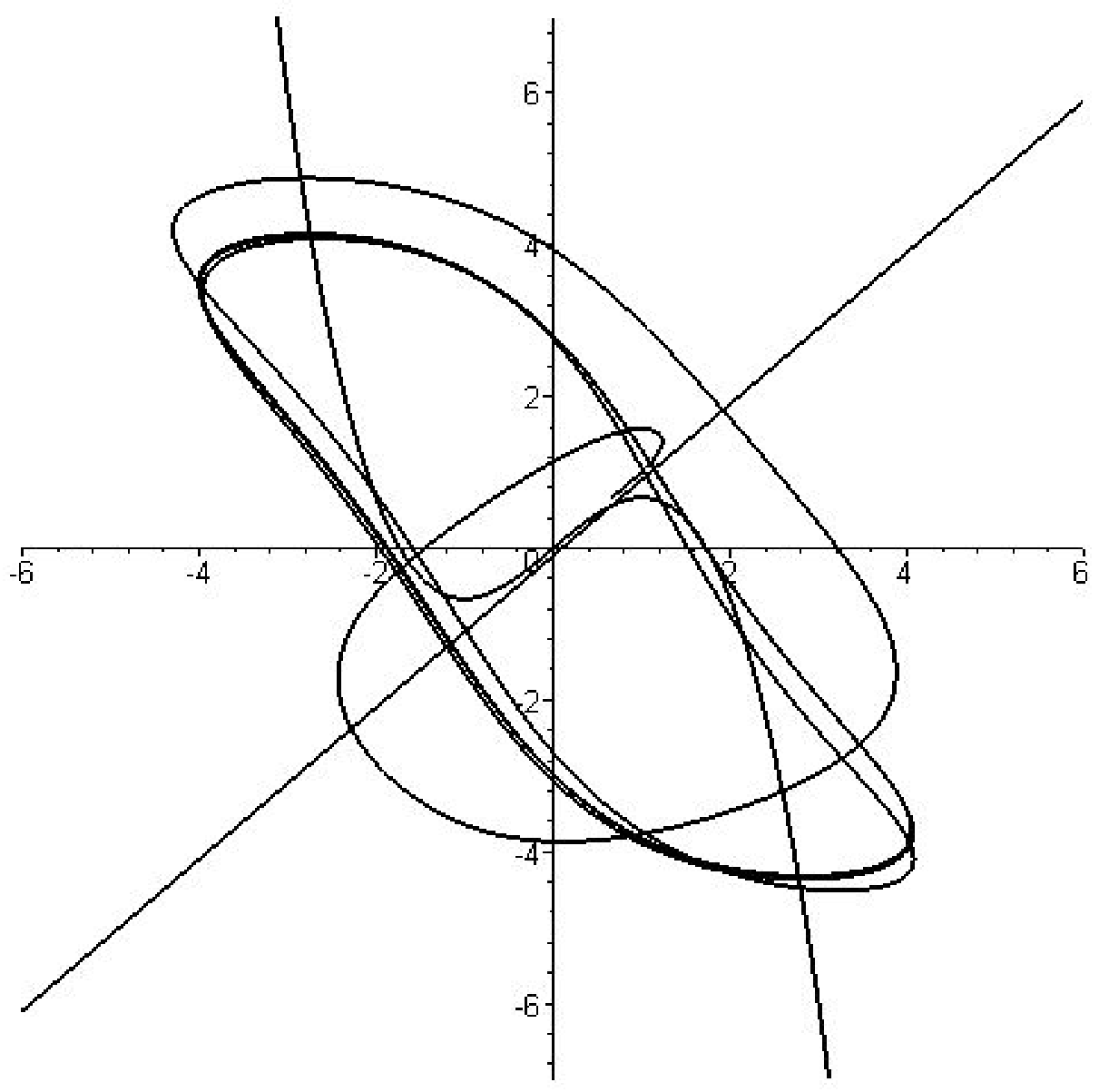} & %
\includegraphics[width=0.35\textwidth]{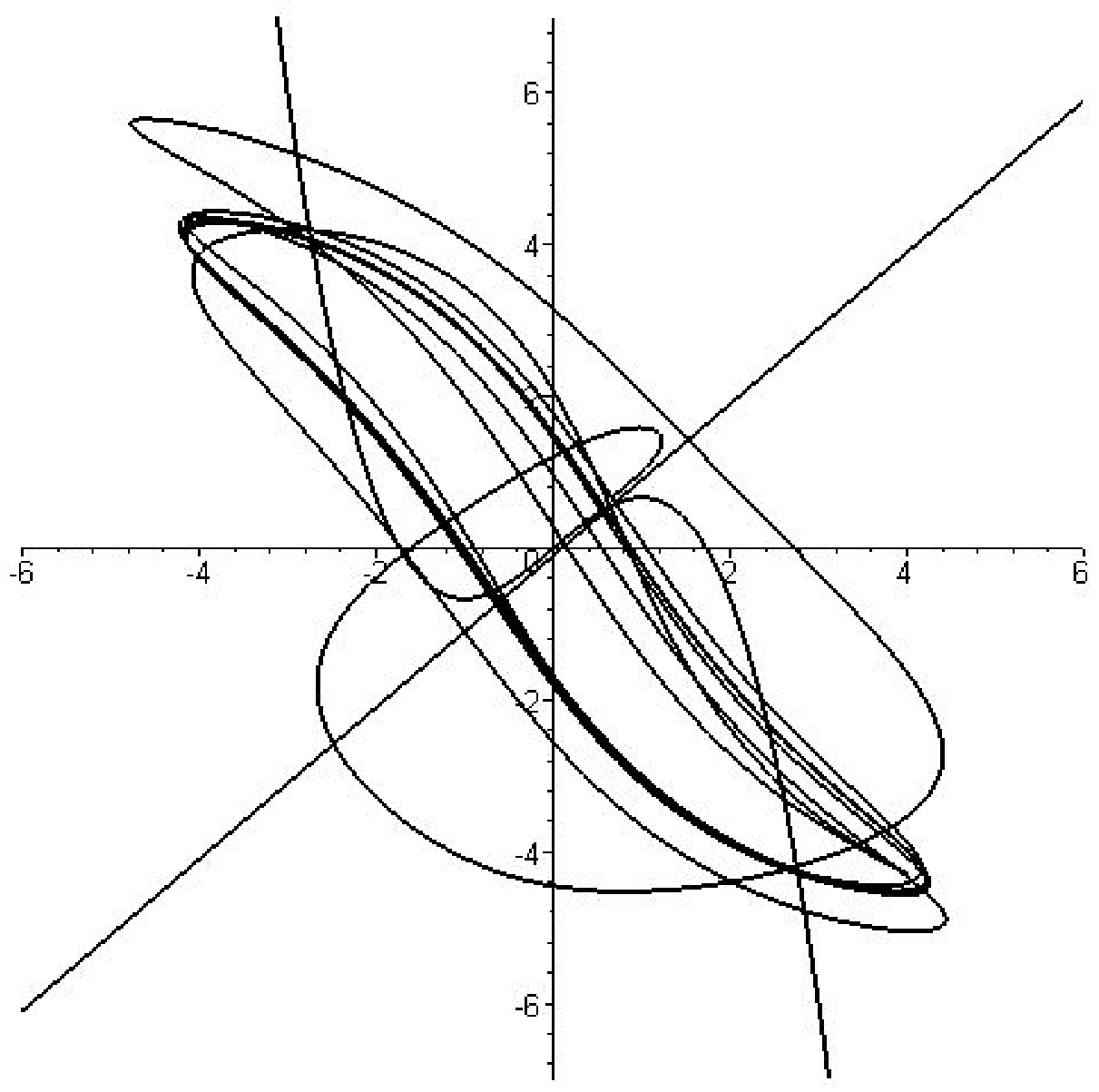} \\
(c) & (d)
\end{tabular}
\begin{tabular}{cc}
\includegraphics[width=0.35\textwidth]{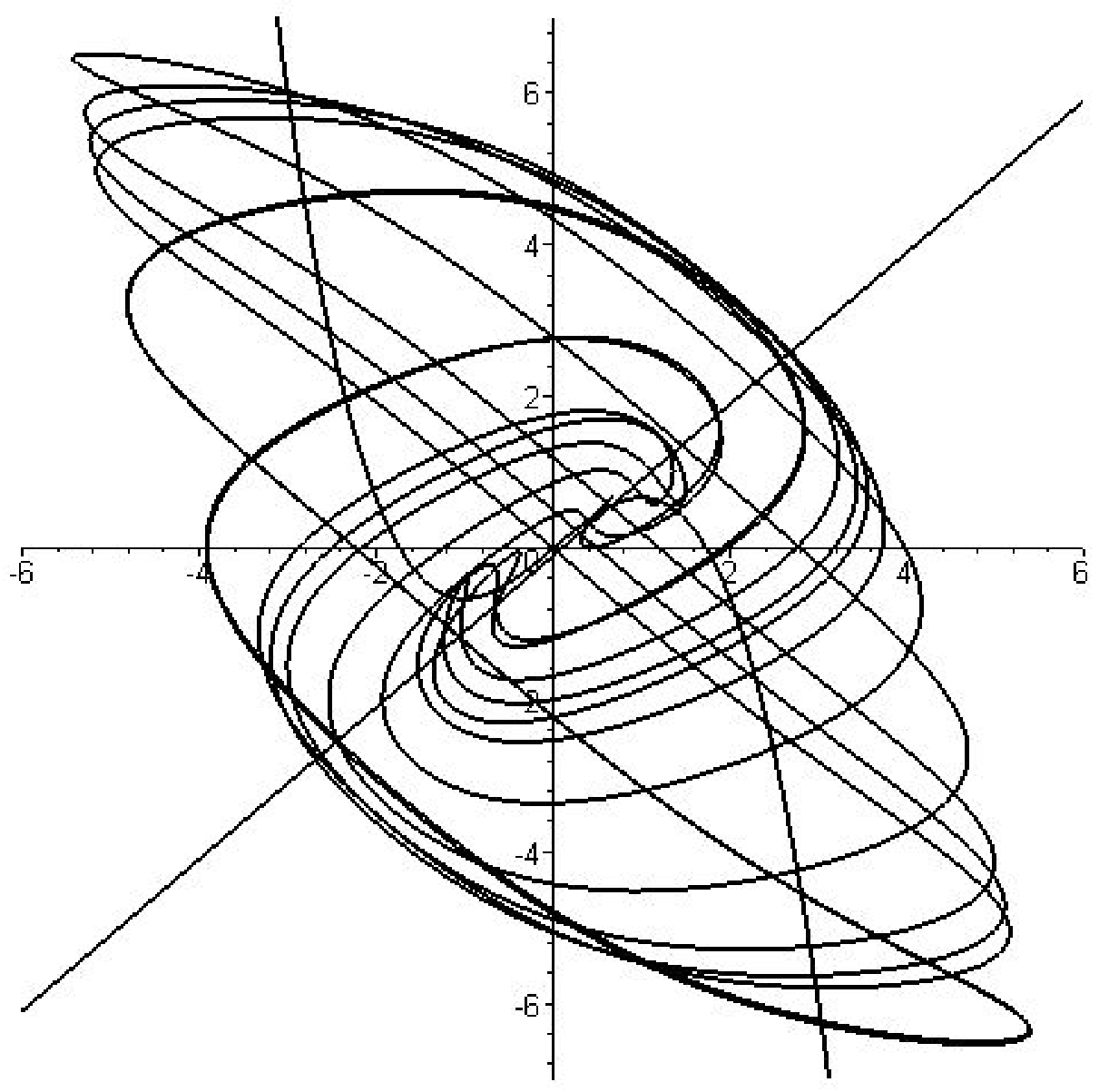} & %
\includegraphics[width=0.35\textwidth]{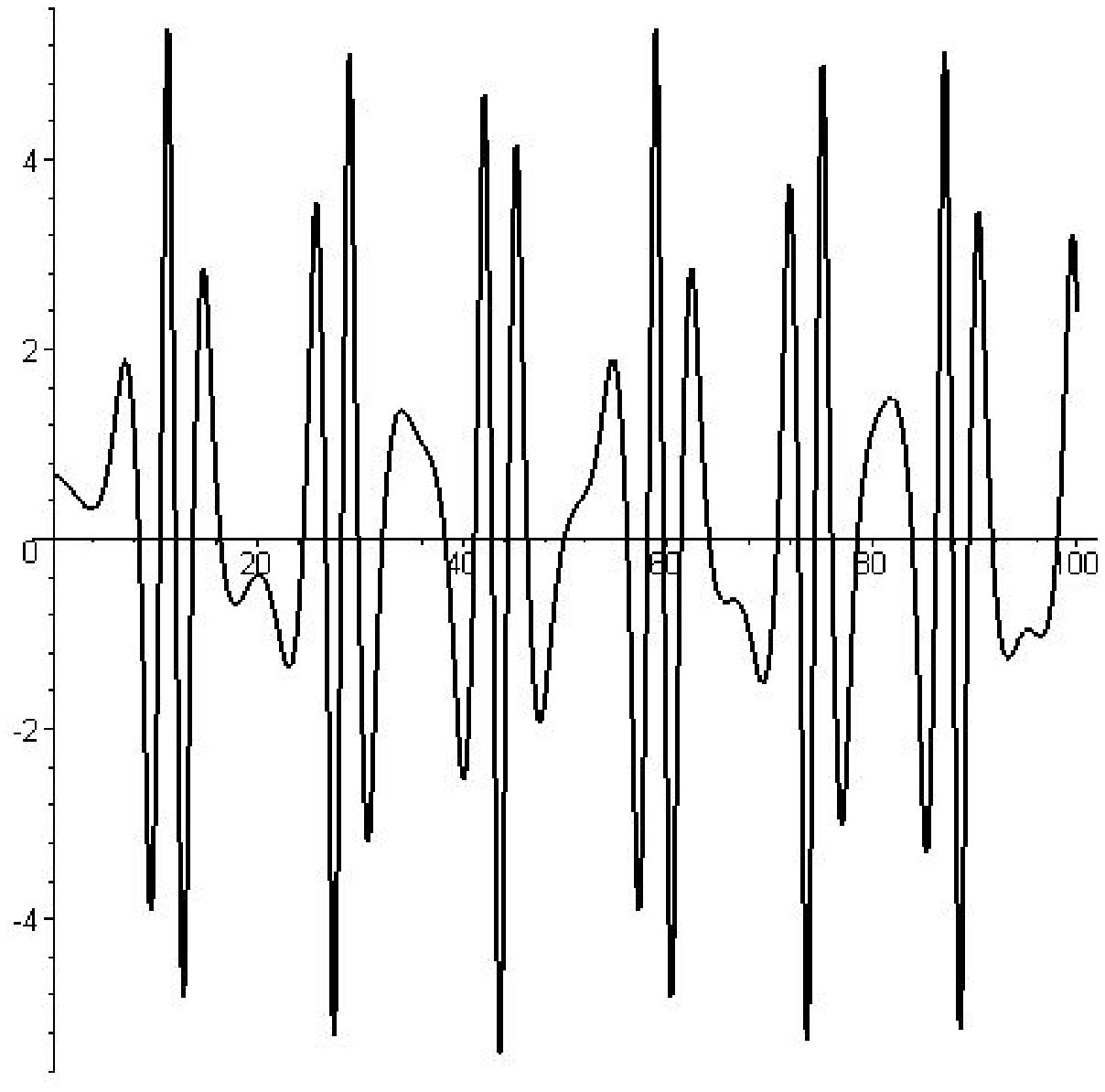} \\
(e) & (f)
\end{tabular}
\end{center}
\caption{Two dimensional phase portrait (a)-(e) and time domain oscillations
corresponding to plot (e) - (f) of the system (\ref{3} ),(\ref{4}) with
nonlinearities (\ref{nlin2}) for $\mathcal{A}=-0.1, \protect\beta=1, \protect%
\tau _{1}=\protect\tau _{2}=1, l=L=0$. \quad (a) -- $\protect\alpha$=1.3,
(b) -- $\protect\alpha $=1.7, (c) -- $\protect\alpha$=1.8, (d) -- $\protect%
\alpha$=1.90, (e) -- $\protect\alpha$=2.0, \quad time domain
oscillations \quad (f) -- $\protect\alpha$=2.00} \label{rys4}
\end{figure}

\begin{figure}[tbp]
\begin{center}
\begin{tabular}{cc}
\includegraphics[width=0.35\textwidth]{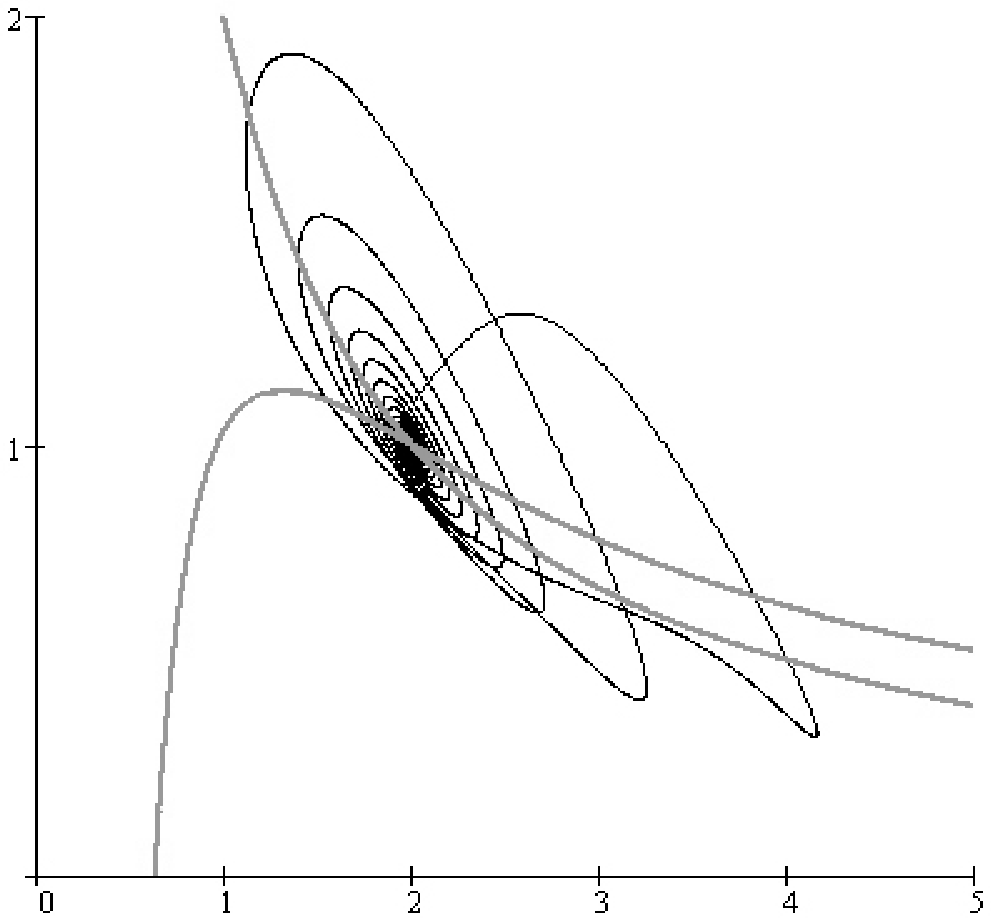} & %
\includegraphics[width=0.35\textwidth]{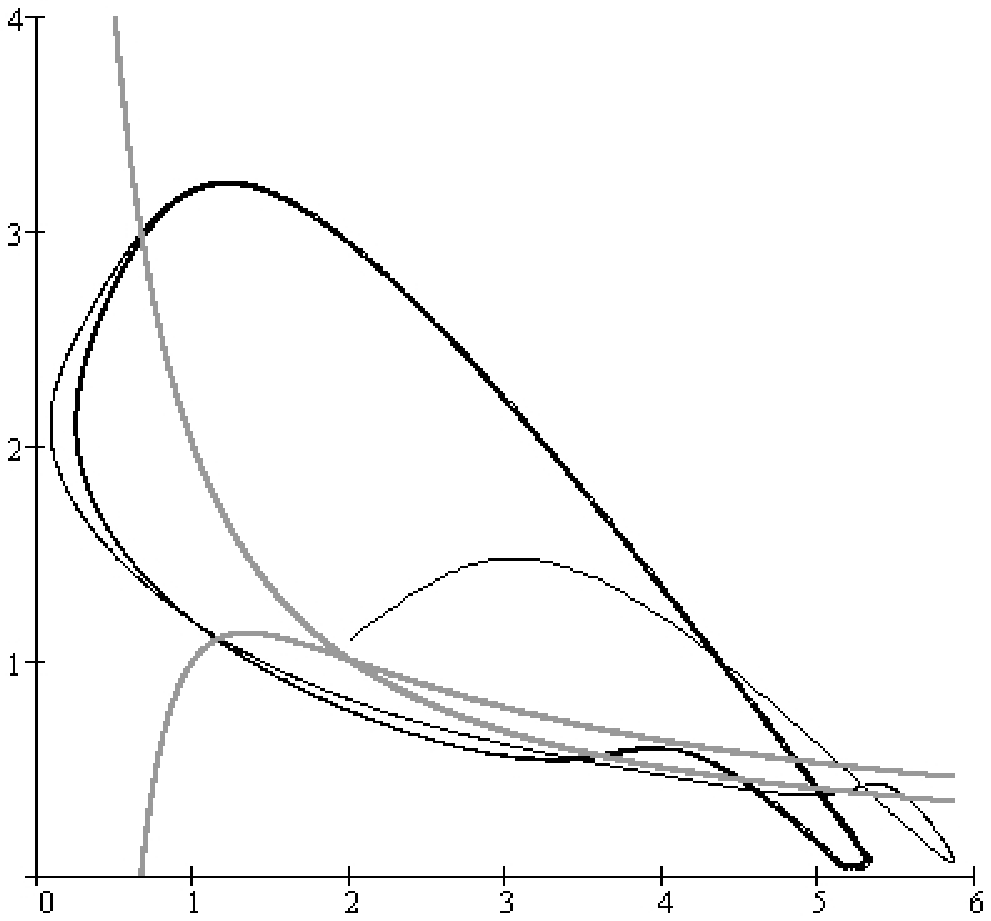} \\
(a) & (b)
\end{tabular}
\begin{tabular}{cc}
\includegraphics[width=0.35\textwidth]{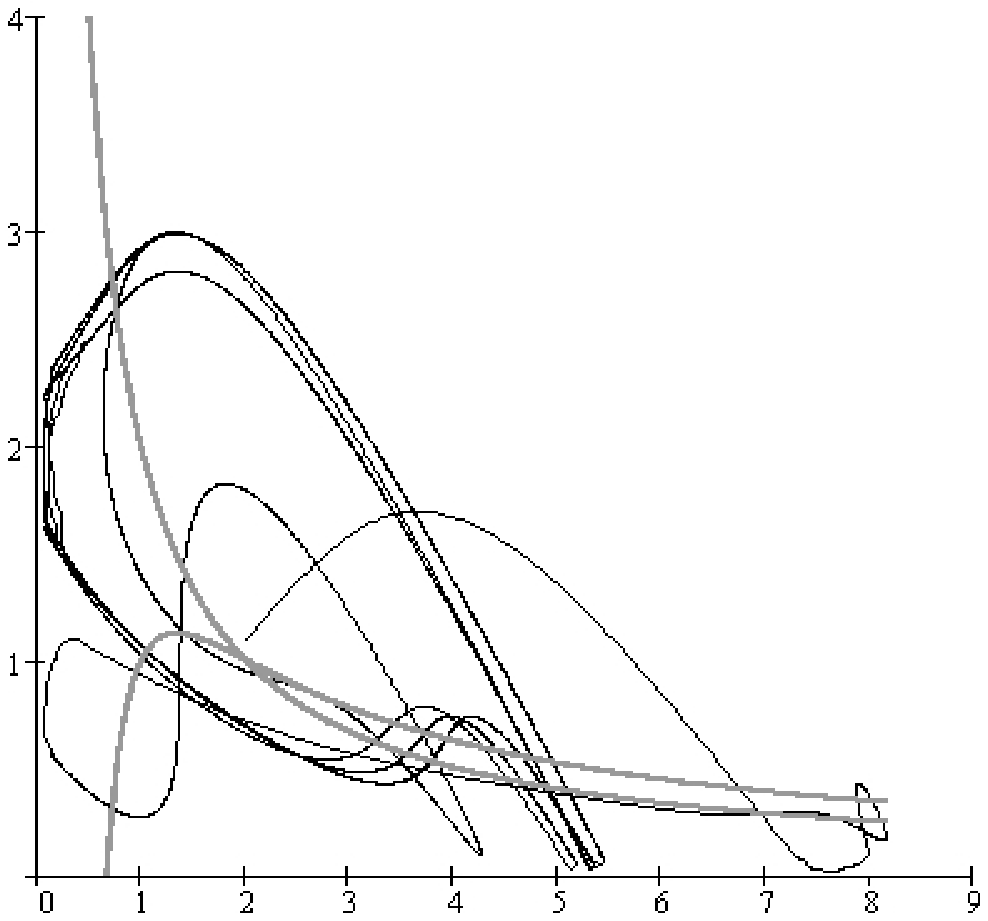} & %
\includegraphics[width=0.35\textwidth]{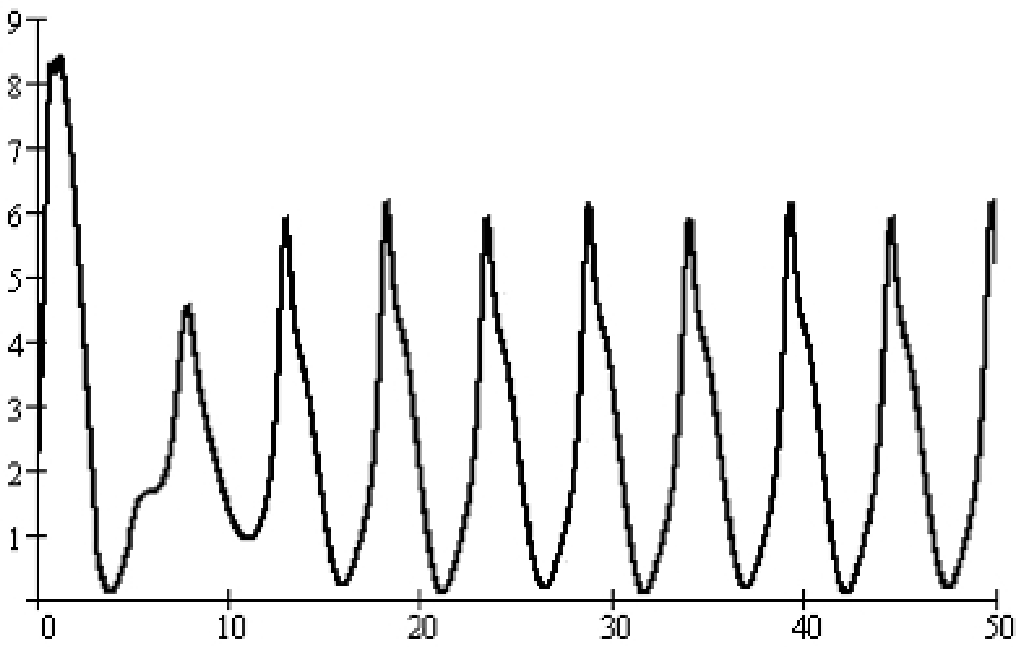} \\
(c) & (d)
\end{tabular}
\end{center}
\caption{Two dimensional phase portrait (a)-(c) and time domain oscillations
corresponding to plot (c) - (d) of the system (\ref{3} ),(\ref{4}) with
nonlinearities (\ref{nlin1}) for $\mathcal{A}=2, \protect\beta=2, \protect%
\tau _{1}=\protect\tau _{2}=1, l=L=0$. \quad (a) --
$\protect\alpha$=1.5, (b) -- $\protect\alpha $=1.6, (c) --
$\protect\alpha$=1.7, \quad time domain oscillations \quad (f) --
$\protect\alpha$=1.7} \label{rys5}

\end{figure}

For $\alpha >\alpha _{0}$ small perturbation of the steady state
solution, due to the memory inherent in fractional derivatives
survive in the process of evolution and grow in amplitude while
nonlinear terms of the system (\ref {nlin2}) restrict the value of
these oscillations. In this case, time dependence of the variables
corresponds to the oscillatory solution. (The phase portrait and
isoclines are presented on Figure \ref{rys4}(a)-(e)).

Analyzing the phase trajectory of the FODEs, we can see that the
amplitude of the oscillations increases with increasing $\alpha.$
At $\alpha $ approaching 2, the oscillations become more
complicated and at $\alpha =2 $ they look more quasichaotic. The
time dependence of this solution is represented in Figure
\ref{rys4}(f).

Brusselator system with nonlinearities (\ref{nlin1}) has quite similar
behavior. Calculating, for example, at $\mathcal{A}=2,  \beta=2, \tau
_{1}=\tau _{2}=1$ the value of $\alpha_{0}$ we find that $\alpha_{0}=1.54$.

The phase portrait and isoclines for (\ref{nlin1}) are presented
in Figure \ref{rys5}
(a)-(c). At $\alpha \lesssim 1.5$, we obtain steady state solution and at $%
\alpha \gtrsim 1.5$ - the steady state oscillation. The increase
of the value $\alpha $ leads to the complication of the phase
paths, and the two-dimensional phase portrait looks much more
complicated (Figure \ref{rys5}(b)-(c)). The attractor of the system of
the two coupled nonlinear differential equations gets the features
of strange attractor and at $\alpha \rightarrow 2 $ it corresponds
to the attractor of the fourth order differential equations
determined by the nonlinearities (\ref{nlin1}).

\section{Computer simulation of pattern formation}

This section contains a discussion of the results of the numerical
study of the initial value problem of the system
(\ref{3})(\ref{4}). The system with corresponding initial and
boundary conditions was integrated numerically using the explicit
and implicit schemes with respect to time and centered difference
approximation for spatial derivatives. The fractional derivatives
were approximated using two different schemes on the basis
Riemann-Liouville definition : $L1$-scheme for $0 \leq \alpha <1$,
$L2$-scheme for $1 \leq \alpha < 2$ (see below and \cite{oldh}),
as well as the scheme on the basis of Grunwald-Letnikov definition
for $0 < \alpha < 1$ and $1 < \alpha < 2$ \cite{po} . In other
words, for the system of $n$ fractional RD equations
\begin{equation} \tau_j \frac{^C\partial^{\alpha_j}
u_j(x,t)}{\partial t^{\alpha_j}}=d_j\frac{\partial^2
u_j(x,t)}{\partial x^2}+f_j(u_1,...,u_n), \qquad j=\overline{1,n},
\end{equation}
where $\tau_j, d_j, f_j $ -- certain parameters and nonlinearities of
the RD system correspondingly, we used the next numerical schemes:
\newline
\bigskip
{\bf L1-scheme}
$$
\delta_j u_{j,i}^{k}-\frac{d_j}{(\Delta
x)^2}(u_{j,i-1}^{k}-2u_{j,i}^{k}+u_{j,i+1}^{k})-
f_j(u_{1,i}^{k},...,u_{n,i}^{k})=
$$
$$
=-\delta_j \left(u_{j,i}^{0}w_k^{(\alpha_j)}+
\sum\limits_{l=1}^{k-1}u_{j,i}^{l}\beta_{k-l+1}^{(\alpha_j)}\right)+\tau_j\frac{(k
\Delta t)^{-\alpha_j}}{\Gamma(1-\alpha_j)}u_{j,i}^{0}
$$
$$
\delta_j=\frac{\tau_j (\Delta t)^{-\alpha_j}}{\Gamma(2-\alpha_j)},
\quad
w_k^{(\alpha_j)}=\frac{1-\alpha_j}{k^{\alpha_j}}-k^{1-\alpha_j}-(k-1)^{1-\alpha_j},
$$
$$
\beta_{s}^{(\alpha_j)}=s^{1-\alpha_j}-2(s-1)^{1-\alpha_j}+(s-2)^{1-\alpha_j},
\quad s=\overline{2,k};
$$
\bigskip
{\bf L2-scheme}
$$
\delta_j u_{j,i}^{k+1}- \frac{d_j}{\Delta
x^2}(u_{j,i-1}^{k+1}-2u_{j,i}^{k+1}+u_{j,i+1}^{k+1})-
f_j(u_{1,i}^{k+1},...,u_{n,i}^{k+1})=
$$
$$
=-\delta_j [ u_{j,i}^{0}w_{0,k}^{\alpha_j} +
u_{j,i}^{1}w_{1,k}^{\alpha_j}+\sum\limits_{l=2}^{k-1}u_{j,i}^{l}\beta_{k-l+2}^{(\alpha_j)}
+ u_{j,i}^{k}(2^{2-\alpha_j}-3) ] + \tau_j \sum \limits_{p=0}^{m}
\frac{(k \Delta t)^{p-\alpha_j}} {\Gamma(p-\alpha_j+1)}
\frac{\partial^p}{\partial t^p} u_{j,i}^{0},
$$
$$
\delta_j=\frac{\tau_j (\Delta t)^{-\alpha_j}}{\Gamma(3-\alpha_j)},
\quad
w_{0,k}^{\alpha_j}=\frac{(1-\alpha_j)(2-\alpha_j)}{k^{\alpha_j}} -
\frac{(2-\alpha_j)}{k^{\alpha_j-1}} +
k^{2-\alpha_j}-(k-1)^{2-\alpha_j},
$$
$$
w_{1,k}^{\alpha_j}=\frac{(2-\alpha_j)}{k^{\alpha_j-1}}-2k^{2-\alpha_j}+3(k-1)^{2-\alpha_j}-(k-2)^{2-\alpha_j},
$$
$$
\beta_{s}^{(\alpha_j)}=s^{2-\alpha_j}-3(s-1)^{2-\alpha_j}+3(s-2)^{2-\alpha_j}-(s-3)^{2-\alpha_j},
\quad s=\overline{3,k}
$$
\bigskip
\newline and {\bf G-L scheme}
$$
u_{j,i}^{k}-\frac{d_j (\Delta t)^{\alpha_j}}{\tau_j (\Delta x)^2}
\left(u_{j,i-1}^{k}-2u_{j,i}^{k}+u_{j,i+1}^{k}
\right)-\frac{(\Delta
t)^{\alpha_j}}{\tau_j}f_j(u_{1,i}^{k},...,u_{n,i}^{k})=
$$
$$
=(\Delta t)^{\alpha_j}\sum \limits_{p=0}^{m}\frac{(k \Delta
t)^{p-\alpha_j}}{\Gamma(p-\alpha_j+1)}\frac{\partial^p}{\partial
t^p} u_{j,i}^{0}-\sum \limits_{l=1}^{k}
c_{l}^{(\alpha_j)}u_{j,i}^{k-l},
$$
$$
c_{0}^{(\alpha_j)}=1, \qquad
c_{l}^{(\alpha_j)}=c_{l-1}^{(\alpha_j)}\left(1-\frac{1+\alpha_j}{l}\right),
\qquad l=1,2,...
$$
where $u_{j,i}^{k} \equiv u_j(x_i,t_k) \equiv u_j(i \Delta x, k
\Delta t), \quad m=[\alpha]$.

The applied numerical schemes are implicit, and for each time layer they are
presented as the system of algebraic equations solved by Newton-Raphson
technique. Such approach makes it possible to get the system of equations
with band Jacobian for each node and to use the sweep method for the
solution of linear algebraic equations. Calculating the values of the
spatial derivatives and corresponding nonlinear terms on the previous layer,
we obtained explicit schemes for integration. Despite the fact that these
algorithms are quite simple, they are very sensitive and require small steps
of integration, and they often do not allow to find numerical results. In
contrast, the implicit schemes, in certain sense, are similar to the
implicit Euler's method, and they have shown very good behavior at the
modeling of fractional reaction-diffusion systems for different step size of
integration, as well as for nonlinear function and the power function of
fractional index. Moreover, by modeling according to this algorithm system (%
\ref{1}),(\ref{2}), we have observed that these results fully match the
results obtained prior.

It should be noted that the definition of the fractional derivative in
Grunwald-Letnikow form is equivalent to the one in Riemann-Liouville method,
but for numerical calculations it is much more flexible.

\begin{figure}[tbp]
\begin{center}
\begin{tabular}{cc}
\includegraphics[width=0.5%
\textwidth]{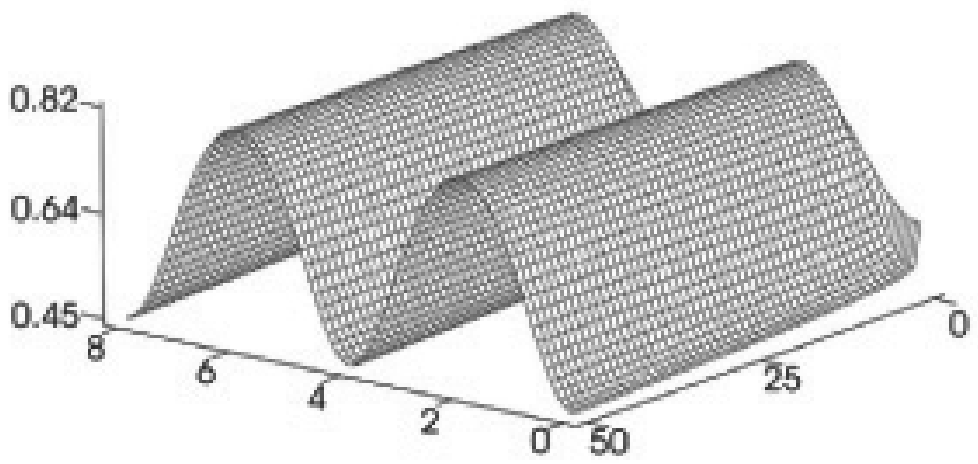} & %
\includegraphics[width=0.5%
\textwidth]{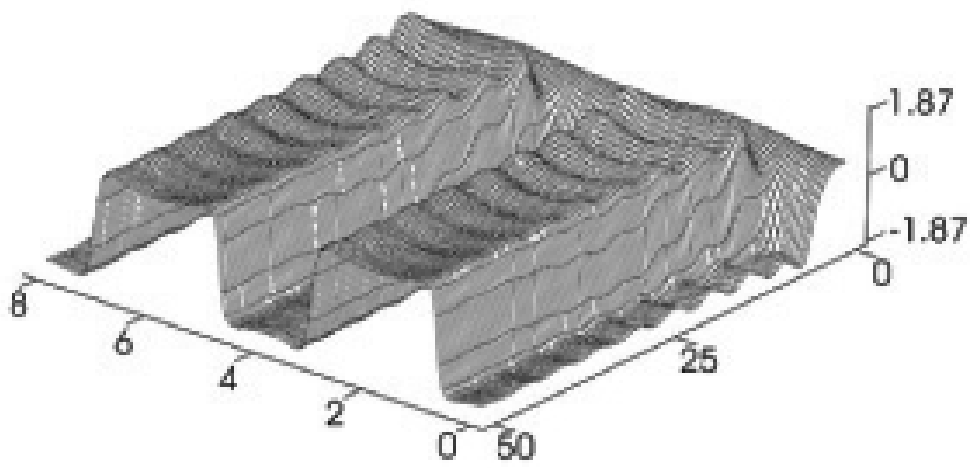} \\
(a) & (d)
\end{tabular}
\begin{tabular}{cc}
\includegraphics[width=0.5%
\textwidth]{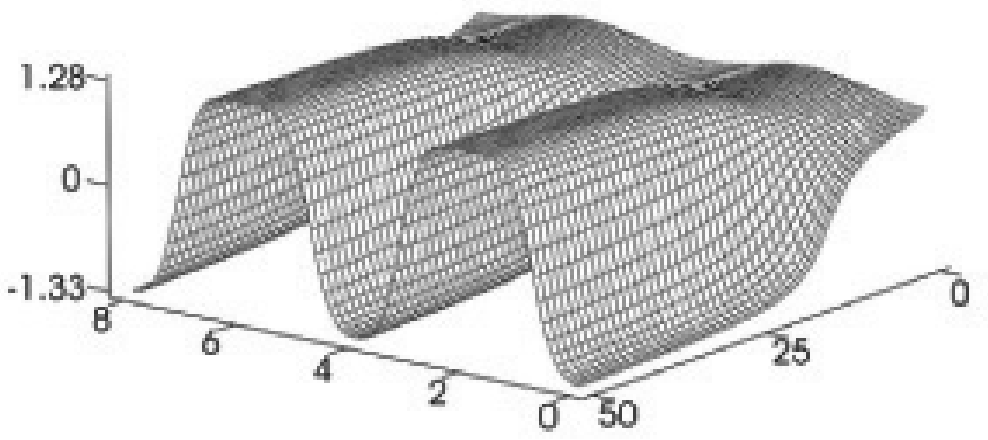} & %
\includegraphics[width=0.5%
\textwidth]{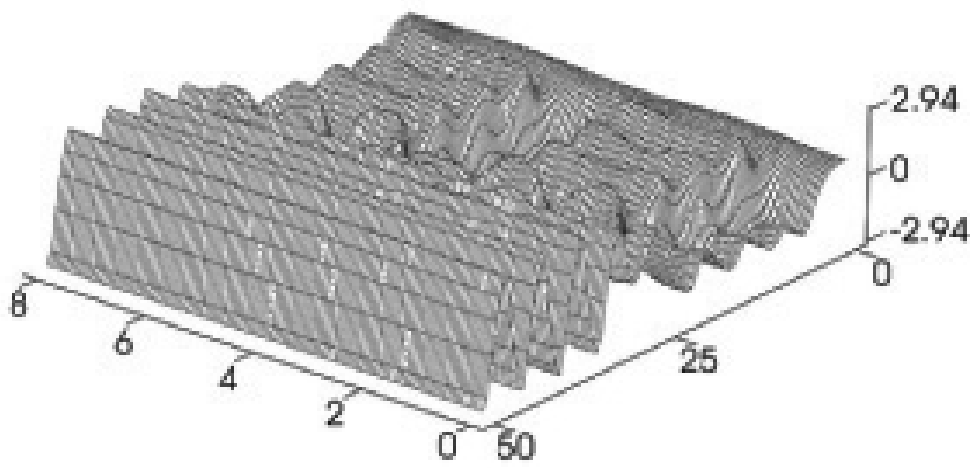} \\
(b) & (e)
\end{tabular}
\begin{tabular}{cc}
\includegraphics[width=0.5%
\textwidth]{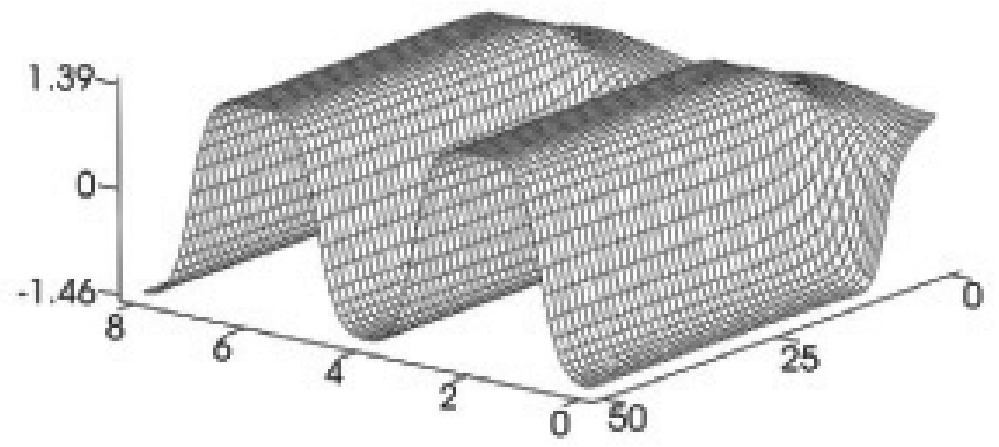} & %
\includegraphics[width=0.5%
\textwidth]{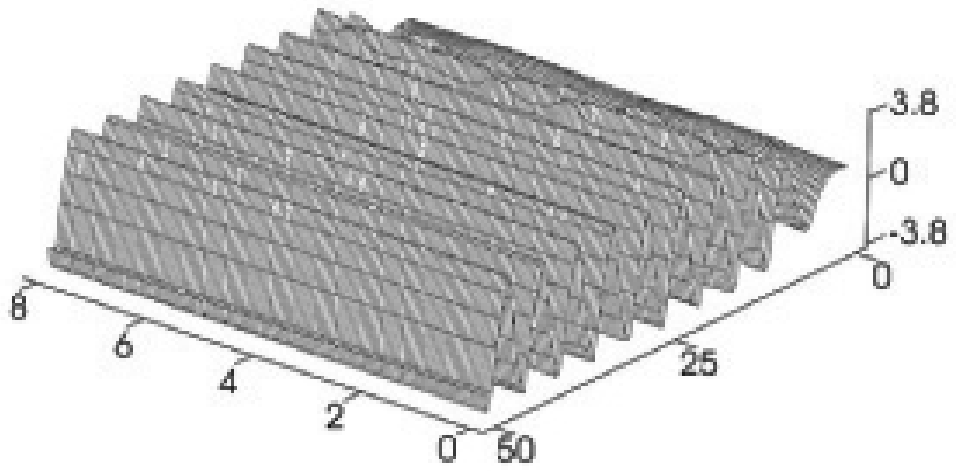} \\
(c) & (f)
\end{tabular}
\end{center}
\caption{Numerical solution of the fractional reaction-diffusion equations (%
\ref{3} ),(\ref{4}) with nonlinearities (\ref{nlin2}).Dynamics of variable $%
n_1$ on the time interval (0,50) for $l_x=8, \mathcal{A}=-0.1, \protect\beta%
=1, \protect\tau _{1}= \protect\tau _{2}=1, \quad l^{2}=0.05, L^{2}=1$; (a) --
$\protect\alpha$=0.1, (b) -- $\protect\alpha$=0.5, (c) -- $\protect\alpha$=0.99,
(d) -- $\protect\alpha$=1.5, (e) -- $\protect\alpha$= 1.6, (f) -- $\protect\alpha$%
=1.8} \label{rys6}

\end{figure}
\begin{figure}[tbp]
\begin{center}
\begin{tabular}{cc}
\includegraphics[width=0.5%
\textwidth]{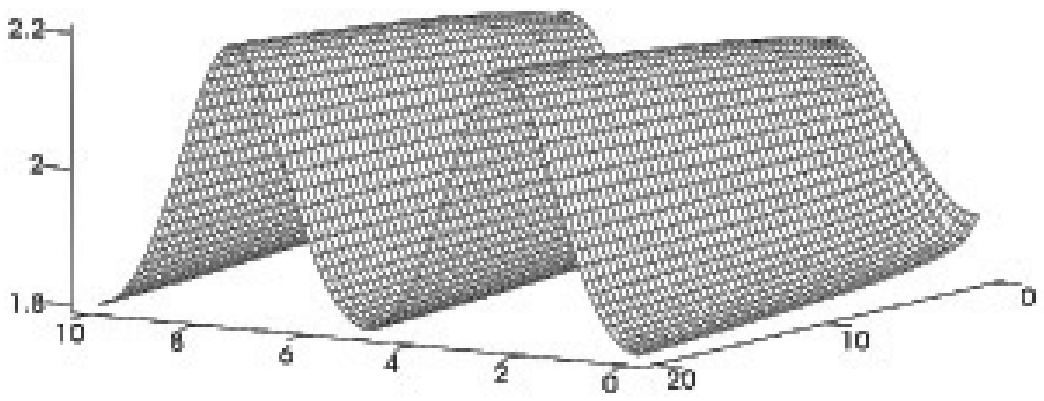} & %
\includegraphics[width=0.5%
\textwidth]{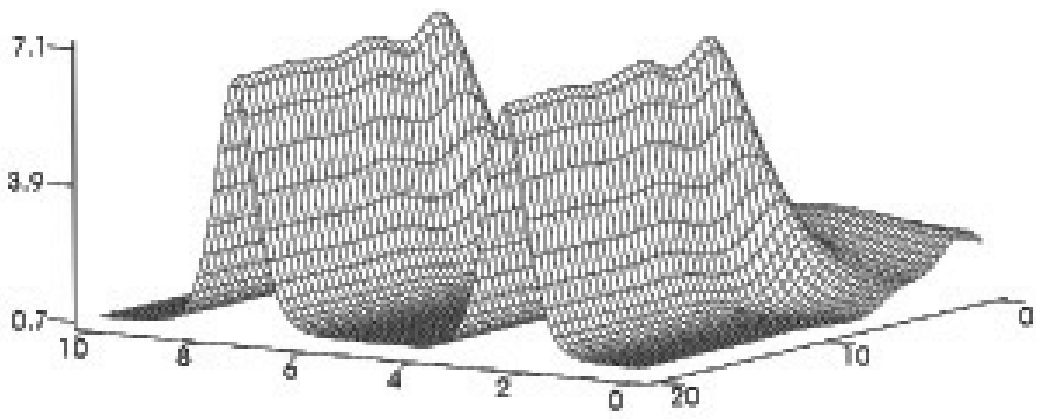} \\
(a) & (d)
\end{tabular}
\begin{tabular}{cc}
\includegraphics[width=0.5%
\textwidth]{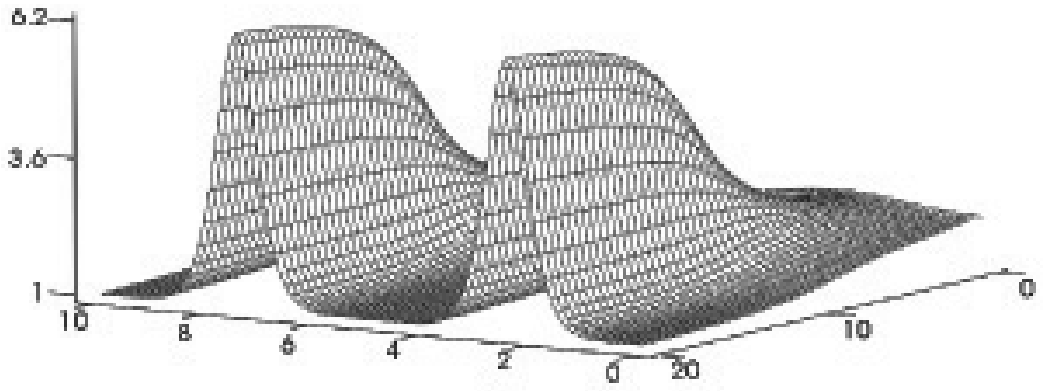} & %
\includegraphics[width=0.5%
\textwidth]{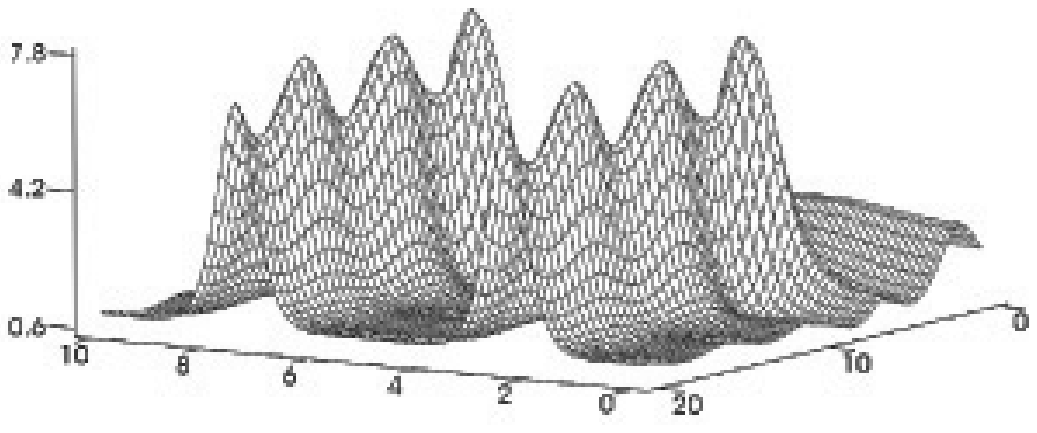} \\
(b) & (e)
\end{tabular}
\begin{tabular}{cc}
\includegraphics[width=0.5%
\textwidth]{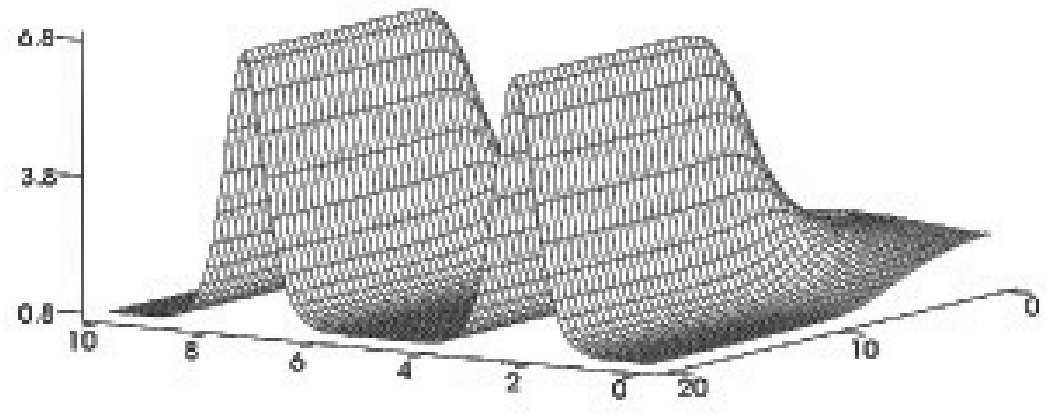} & %
\includegraphics[width=0.5%
\textwidth]{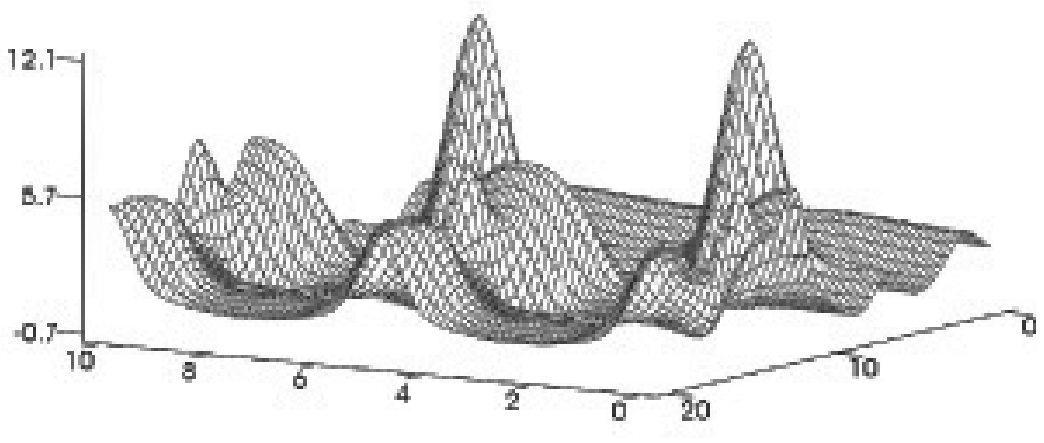} \\
(c) & (f)
\end{tabular}
\end{center}
\caption{Numerical solution of the fractional reaction-diffusion equations (%
\ref{3} ),(\ref{4}) with nonlinearities (\ref{nlin1}).Dynamics of variable $%
n_1$ on the time interval (0,20) for $l_x=10, \mathcal{A}=2, \protect\beta%
=2, \protect\tau _{1}= \protect\tau _{2}=1,\quad l^{2}=0.1, L^{2}=10 $; (a) --
$\protect\alpha$=0.1, (b) -- $\protect\alpha$=0.5, (c) -- $\protect\alpha$=0.99,
(d) -- $\protect\alpha$=1.5, (e) -- $\protect\alpha$= 1.6, (f) -- $\protect\alpha$%
=1.8} \label{rys7}

\end{figure}

We have considered here the kinetics of formation of dissipative
structures for different values of $\alpha $. These results are
presented on Figures \ref{rys6} and \ref{rys7}.

The simulations were carried out for a one-dimensional system on an
equidistant grid with spatial step $h$ changing from = 0.1 to 0.01 and time
step $\Delta t$ changing from 0.001 to 0.1. We used imposed Neuman (\ref
{bc1})\ or periodic boundary conditions (\ref{bc2}). As the initial
condition, we used the uniform state which was superposed with a small
spatially inhomogeneous perturbation.

The systems have rich dynamics, including steady state dissipative
structures, homogeneous and nonhomogeneous oscillations, and spatiotemporal
patterns. In this paper, we focus mainly on the study of general properties
of the solutions depending on the value of $\alpha$.

As discussed in Section 2, there are two different regions in
parameter $\mathcal{A}$, where the system can be stable or unstable. In the
case of $\alpha =1$ the steady state solutions in the form of
nonhomogeneous dissipative structures are inherent to unstable
region $\overline{n}_{1}\in (-1,1)$. Figures \ref{rys6}(a)-(c) show
the steady state dissipative structure formation and Figures
\ref{rys6} (d)-(f) present the spatio-temporal evolution of
dissipative structures, which eventually leads to homogeneous
oscillations.

On the Figure \ref{rys6}(a)-(d), the value $\alpha $ increases from \ $0.1$
to $1.5 $ and on this whole interval the structures are in steady
state. This is due to the case $\alpha <\alpha _{0}$, the
oscillatory perturbations are damping, and we can see that small
oscillations are at the transition period $n_{1}.$ With increasing
$\alpha $, the steady state structures change to the
spatio-temporary behavior (Figure \ref{rys6}(e)-(f)).

The emergence of homogeneous oscillations, which destroy pattern
formation (Figure \ref{rys6}(e),(f))\ \, has deep physical meaning. The
matter is that the stationary dissipative structures consist of
smooth and sharp regions of variable $n_{1}$, and the smooth shape
of $n_{2}$. The linear system analysis shows that the homogeneous
distribution of the variables is
unstable according to oscillatory perturbations inside the wide interval of $%
\overline{n}_{1},$ which is much wider then interval ($-1,1$). At the same
time, smooth distributions at the maximum and minimum values of $n_{1}$ are $%
\pm \sqrt{3}$ correspondingly. In the first approximation, these smooth
regions of the dissipative structures resemble homogeneous ones and are
located inside the instability regions. As a result, the unstable
fluctuations lead to homogeneous oscillations, and the dissipative
structures destroy themselves. We can conclude that oscillatory modes in
such type FODEs have a much wider attraction region than the corresponding
region of the dissipative structures.

For a wide range of the parameters ${\alpha }$, the numerical
solutions of the Brusselator problem show similar behavior (Figure
\ref{rys7}(a)-(d)). The stationary solutions emerge practically in
the same way. At small ${\alpha }$, we see aperiodic formation of
the structures, and approaching ${\alpha }_{0}$, the
damping oscillations of the dissipative structures arise. At ${\alpha }%
_{0}=1.7$ certain non-stationary structures arise (Figure
\ref{rys7}(e)). In this case, the dissipative structures look
quite similar to those we obtained for regular system
\cite{gk91,gd98}. The increase of $\alpha $ leads to a larger
amplitude of pulsation. All these patterns are robust with respect
to small initial perturbations. The further increase of $\alpha $
leads to spatially temporary chaos (Figure \ref{rys7}(f)).

In the contrast to previous case, such nonhomogeneous behavior is stable and
does not lead to homogeneous oscillations. The matter is that in Brusselator
model, the dissipative structures are much grater in amplitude and do not
have smooth distribution at the top.

It should be noted that the pulsation phenomena of the dissipative
structures is closely related to the oscillation solutions of the
ODE (Figures \ref{rys4}, \ref{rys5}). Moreover, the fractional
derivative of the first variable has the most impact on the
oscillations emergence. It can be obtained by performing a
simulation where the first variable is a fractional derivative and
the second one is an integer. It should be emphasized that the
distribution of $n_{2}$, within the solution, only shows a small
deviation from the stationary value (that is why this variable is
not represented in the figures).

\section{Conclusion}

In this article we developed a linear theory of instability of reaction
diffusion system with fractional derivatives. The introduced new parameter
-- marginal value ${\alpha }_{0}$ plays the role of bifurcation parameter.
If the fractional derivative index $\alpha$ is smaller than ${\alpha }_{0}$,
the system of FODEs is stable and has oscillatory damping solutions. At $%
\alpha >{\alpha }_{0}$, the FODEs becomes unstable, and we obtain
oscillatory or even more complex - quasi chaotic solutions. In addition, the
stable and unstable domains of the system were investigated.

By the computer simulation of the fractional reaction-diffusion systems we
provided evidence that pattern formation in the fractional case, at $\alpha $
less than a certain value, is practicably the same as in the regular case
scenario $\alpha =1$. At $\alpha>{\alpha }_{0}$, the kinetics of formation
becomes oscillatory. At $\alpha ={\alpha }_{0}$, the oscillatory mode arises
and can lead to homogeneous or nonhomogeneous oscillations. In the last case
scenario, depending on the parameters of the medium, we can see a rich
variety of spatiotemporal behavior.


\begin{thebibliography}{99}
\bibitem{pr}  \textsc{G. Nicolis, I. Prigogine}. Self-organization in
non-equilibrium systems. Wiley, New York. 1977.

\bibitem{ch}  \textsc{M. C. Cross and P. S. Hohenberg.} \emph{Pattern
formation outside of equilibrium}, Rev. Modern Phys., 65 (1993), pp.
851--1112.

\bibitem{m90}  \textsc{A. S. Mikhailov.} \emph{Foundations of Synergetics},
Springer-Verlag, Berlin, 1990.

\bibitem{KO}  \textsc{B.S. Kerner, V.V. Osipov.} \emph{Autosolitons} Kluwer,
Dordrecht, 1994.

\bibitem{dk89}  \textsc{J. D. Dockery and J. P. Keener} \emph{Diffusive
effects on dispersion in excitable media}, SIAM J. Appl. Math., 49 (1989),
pp. 539--566.

\bibitem{dk03}  \textsc{A. Doelman and T. J. Kaper.} \emph{Semistrong pulse
interactions in a class of coupled reaction-diffusion equations}
SIAM J. Applied dynamical systems Vol. 2, No. 1, (2003), pp.
53--96.

\bibitem{lg}  \textsc{A.Lubashevskii, V.V.Gafiychuk.} \emph{The projection
dynamics of highly dissipative system}. Phys. Rev. E. vol.50, No.1,
(1994), pp.171--181.

\bibitem{gl}  \textsc{V.V. Gafiychuk, I.A. Lubashevskii.} \emph{Variational
representation of the projection dynamics and random motion of highly
dissipative systems.} J. Math. Phys. v.36. \#10, (1995), pp. 5735-5752.

\bibitem{sbg}  \textsc{P.Schutz, M.Bode, V.V.Gafiychuk.} \emph{Transition
from stationary to travelling localized patterns in two-dimensional
reaction-diffusion system. }Phys. Rev. E. v.52., N4, (1995), pp. 4465-4473.

\bibitem{mo02}  \textsc{C. B. Muratov and V. V. Osipov.} \emph{Stability of
the static spike autosolitons in the Gray-scott model} Siam J. Appl. Math.
Vol. 62, no. 5, (2002), pp. 1463--1487.

\bibitem{g99}  \textsc{M. Golubitsky, D. Luss and S.H. Strogatz}. \emph{Pattern Formation in
Continuous and Coupled Systems}, IMA Volumes in Mathematics and
its Applications 115 , Springer, New York, (1999).

\bibitem{hw}  \textsc{B.I. Henry, T.A.M. Langlands and S.L. Wearne } \emph{%
Turing pattern formation in fractional activator-inhibitor
systems}. Phys. Rev. E., 72, \# 026101, (2005).

\bibitem{hw1}  \textsc{B.I. Henry, S.L. Wearne.} \emph{Fractional
reaction-diffusion.} Physica A 276, (2000), pp. 448--455.

\bibitem{he02}  \textsc{B.I. Henry and S.L. Wearne.} \emph{Existence of
turing instabilities in a two-species fractional
reaction-diffusion system}. Siam J. Appl. Math. Vol. 62, No. 3,
(2002), pp. 870--887.

\bibitem{add}  \textsc{M. O. Vlad and J. Ross.} \emph{Systematic derivation
of reaction-diffusion equations with distributed delays and relations to
fractional reaction-diffusion equations and hyperbolic transport equations:
Application to the theory of Neolithic transition}. Phys. Rev. E 66, 061908
(2002).

\bibitem{add1}  \textsc{K. Seki, M. Wojcik, and M. Tachiya.} \emph{%
Fractional reaction-diffusion equation}. J. Chem. Phys. 119, 2165, (2003).

\bibitem{gd}  \textsc{V. Gafiychuk, B. Datsko.} \emph{Pattern formation in a
fractional reaction-diffusion system}. Physica A: Statistical
Mechanics and its Applications. Vol. 365, (2006), pp. 300--306.

\bibitem{GDI05}  \textsc{V.V.Gafiychuk, B.Y.Datsko,
Yu.Yu.Izmajlova}.   Analysis of the dissipative structures in
reaction-diffusion systems with fractional derivatives. Math.
metody ta phys.-mech. polia. V.49, \#4, 2006, pp. 109-116 (in
Ukrainian)

\bibitem{ar1}  \textsc{R.K. Saxena, A.M. Mathai and H.J. Haubold.}
Fractional reaction-diffusion equations. arXiv:math.CA/0604473 v1 21 Apr
2006.

\bibitem{zw}  \textsc{G.M. Zaslavsky and H. Weitzner} Some Applications of
Fractional Equations. E-print nlin.CD/0212024, (2002).

\bibitem{vb}  \textsc{C. Varea and R. A. Barrio.} Travelling Turing patterns
with anomalous diffusion. J. Phys.: Condens. Matter 16, (2004), pp. 5081--5090.

\bibitem{tz}  \textsc{\ V. E Tarasov and G. M Zaslavsky.} Nonholonomic
constraints with fractional derivatives. J. Phys. A: Math. Gen. 39
(2006), pp. 9797--9815.

\bibitem{skm}  \textsc{S.G. Samko, A. A. Kilbas, and O. I. Marichev,}\emph{\
Fractional Integrals and Derivatives: Theory and Applications,} Gordon and
Breach, Newark, N. J., 1993.

\bibitem{po}  \textsc{I. Podlubny}. \emph{Fractional Differential Equations}%
. Academic Press, 1999.

\bibitem{os06}  \textsc{Z. M. Odibat, N. T. Shawagfeh} \emph{Generalized
Taylor's formula. Applied Mathematics and Computation} xxx (2006) xxx-xxx
(available online at www.sciencedirect.com)

\bibitem{a}  \textsc{R Yu, H. Zhang,} \emph{New function of Mittag-Leffler
type and its application in the fractional diffusion-wave equation,} Chaos,
Solitons and Fractals 30 (2006), pp. 946--955.

\bibitem{a94}  \textsc{G. Adomian}. Solving Frontier Problems of Physics:
The Decomposition Method, Kluwer Academic Publishers, Dordrecht, 1994.

\bibitem{bbi04}  \textsc{\ J. Biazar, E. Babolian, R. Islam} Solution of the
system of ordinary differential equations by Adomian decomposition method,
Appl. Math. Comput. 147 (3) (2004), pp. 713--719.

\bibitem{jd06}  \textsc{H. Jafari, V. Daftardar-Gejji}. \emph{Solving a
system of nonlinear fractional differential equations using Adomian
decomposition}. Journal of Computational and Applied Mathematics 196 (2006), pp. 644 - 651

\bibitem{db04}  \textsc{V. Daftardar-Gejji, A. Babakhani}. Analysis of a
system of fractional differential equations, J. Math. Anal. Appl.
293 (2004), pp. 511--522.

\bibitem{zse05}  \textsc{G.M. Zaslavsky, A.A. Stanislavsky, M. Edelman}
Chaotic and Pseudochaotic Attractors of Perturbed Fractional Oscillator,
arXiv:nlin.CD/0508018.

\bibitem{mi}  \textsc{D. Matignon,} \emph{Stability results for fractional
differential equations with applications to control processing, }%
Computational Eng. in Sys. Appl., Vol. 2, Lille, France 963, 1996.

\bibitem{El96}  \textsc{A.M.A. El-Sayed.} \emph{Fractional
differential-difference equations,} J. Fract. Calc. 10 (1996), pp. 101--106.

\bibitem{wz}  \textsc{H.Weitzner, G.M.Zaslavsky}. \emph{Some applications of
fractional equations}. Communications in Nonlinear Science and Numerical
Simulation 8 (2003), pp. 273--281.

\bibitem{oldh}  \textsc{K.D.Oldham and J.Spanier} \emph{The Fractional
Calculus: Theory and Applications of Differentiation and Integration to
Arbitrary Order, Vol.111 of Mathematics in Science and Engineering }
Academic Press, New York, 1974.

\bibitem{gk91}  \textsc{V.V.Gafiychuk, B.S.Kerner, V.V.Osipov,
T.M.Scherbatchenko}. Formation of pulsating thermal-diffusion autosolitons
and turbulence in a nonequilibrium electron-hole plasma. Sov. Phys. Sem.
(USA). v.25, \#11, (1991) (Translation of Fiz.Tekh. Poluprovodn (USSR).
v.25, No.11, (1991), pp. 1696--1702).

\bibitem{gd98}  \textsc{V.V.Gafiychuk, A.V.Demchuk}. Analysis of the
dissipative structures in Gierer-Meinhardt model. Matematicheskie metody i
physico-mechanicheskie polia. V.40, \#2, 1997, pp.48-53 (in Ukrainian,
English translation in Journal of Mathematical Sciences, v.88, \#4, 1998).

\end{thebibliography}
\end{document}